%% file: main.tex
\documentclass[twocolumn]{svjour3}

\usepackage{hyperref}
\usepackage{booktabs}
\usepackage{float}
\usepackage{environ}
\usepackage{svg}
\usepackage[utf8]{inputenc}
\usepackage{amsmath}
\usepackage{graphicx}
\usepackage{color}
\usepackage{graphviz}
\usepackage{xcolor}

\usepackage{bbm}
\usepackage{amsmath}
\usepackage{amssymb}
\usepackage{dsfont}

\usepackage{caption}
\usepackage{subfig} 		
\usepackage{tikz} 			
\usetikzlibrary{arrows}

\usepackage{multirow}

\usepackage[ruled]{algorithm2e}

\usepackage{balance}

\makeatletter
\pgfdeclareshape{datastore} {
	\inheritsavedanchors[from=rectangle]
	\inheritanchorborder[from=rectangle]
	\inheritanchor[from=rectangle]{center}
	\inheritanchor[from=rectangle]{base}
	\inheritanchor[from=rectangle]{north}
	\inheritanchor[from=rectangle]{north east}
	\inheritanchor[from=rectangle]{east}
	\inheritanchor[from=rectangle]{south east}
	\inheritanchor[from=rectangle]{south}
	\inheritanchor[from=rectangle]{south west}
	\inheritanchor[from=rectangle]{west}
	\inheritanchor[from=rectangle]{north west}
	\backgroundpath{
		\southwest \pgf@xa=\pgf@x \pgf@ya=\pgf@y
		\northeast \pgf@xb=\pgf@x \pgf@yb=\pgf@y
		\pgfpathmoveto{\pgfpoint{\pgf@xa}{\pgf@ya}}
		\pgfpathlineto{\pgfpoint{\pgf@xb}{\pgf@ya}}
		\pgfpathmoveto{\pgfpoint{\pgf@xa}{\pgf@yb}}
		\pgfpathlineto{\pgfpoint{\pgf@xb}{\pgf@yb}}
	}
}

\newcommand{\ie}{i.\,e. }

\begin{document}

\title{DUEF-GA: Data Utility and Privacy Evaluation Framework for Graph Anonymization}


\author{Jordi Casas-Roma}

\institute{
    Jordi Casas-Roma \at 
		Universitat Oberta de Catalunya (UOC)\\
		Internet Interdisciplinary Institute (IN3)\\
		Center for Cybersecurity Research of Catalonia (CYBERCAT)\\
		Barcelona, Spain\\
		\email{jcasasr@uoc.edu}
}

\date{Received: date / Accepted: date}

\maketitle 

\begin{abstract}
Anonymization of graph-based data is a problem which has been widely studied over the last years and several anonymization methods have been developed. Information loss measures have been used to evaluate data utility and information loss in the anonymized graphs. However, there is no consensus about how to evaluate data utility and information loss in privacy-preserving and anonymization scenarios, where the anonymous datasets were perturbed to hinder re-identification processes. Authors use diverse metrics to evaluate data utility and, consequently, it is complex to compare different methods or algorithms in literature. In this paper we propose a framework to evaluate and compare anonymous datasets in a common way, providing an objective score to clearly compare methods and algorithms. Our framework includes metrics based on generic information loss measures, such as average distance or betweenness centrality, and also task-specific information loss measures, such as community detection or information flow. Additionally, we provide some metrics to examine re-identification and risk assessment. We demonstrate that our framework could help researchers and practitioners to select the best parametrization and/or algorithm to reduce information loss and maximize data utility.

\end{abstract}

\input{1-intro}
\input{2-concepts}
\input{3-related-work}
\input{4-framework}
\input{5-GIL}
\input{6-SIL}

\input{7-reidentification}
\input{8-examples}
\input{9-conclusions}


\paragraph{Compliance with Ethical Standards.} This work was supported by the Spanish Government, in part under Grant RTI2018-095094-B-C22 ``CONSENT'', and in part under Grant TIN2014-57364-C2-2-R ``SMARTGLACIS''.

\paragraph{Conflict of Interest.} Author Jordi Casas-Roma declares that he has no conflict of interest.

\paragraph{Ethical approval.} This article does not contain any studies with human participants or animals performed by any of the authors.

\balance

\end{document}

%% file: 1-intro.tex
\section{Introduction}
\label{sec:intro}

In recent years, a high number of social and human interaction graphs have been made publicly available. Embedded within this data, there is user's private information that must be preserved before releasing the data to third parties and researchers. The study by Ferri et al. \cite{FerriEtAl:2012} reveals that up to $90\%$ of user groups are concerned about data owners sharing data about them. Backstrom et. al. \cite{BackstromEtAl:2007} point out that the simple technique of anonymizing graphs by removing the identities of the vertices before publishing the actual graph does not always guarantee privacy.

Several anonymization methods have been developed to mitigate this problem. The main objective of an anonymization process is twofold: to preserve the user's privacy and hinder re-identification processes, and also to maintain data utility on anonymized data, \ie minimizing information loss. However, these methods introduce noise in the original data, which may reduce its usefulness in subsequent data mining processes, such as clustering or information flow. Anonymization methods should allow the analysis performed in the anonymized data to lead to results as equal as possible to the ones obtained when applying the same analysis to the original data. Nevertheless, data modification is contrary to data utility. The larger data modification, the less data utility. Thus, a good anonymization method hinders the re-identification process while causing minimal distortion to the data.

Although several metrics have been designed to evaluate the goodness of the anonymization methods, there is no standard or common way to evaluate data utility or information loss. Usually, several authors use different information loss measures to quantify perturbation on anonymized data. Hence, it is very hard (or even impossible) to compare information loss and data utility among different methods and algorithms, since each work of the literature uses specific (and usually different) metrics to evaluate the perturbation on anonymous graphs. 

Our framework aims to help researchers and practitioners to choose the correct parametrization or method to anonymize a graph-based dataset. Some possible scenarios, but not limited to, could be to choose the right anonymization percentage when dealing with random-based anonymization algorithms in order to get a good trade-off between data utility and privacy. Another one could be to compare different $k$-anonymity algorithms to decide the one that better preserves the data utility.

To the best of our knowledge, there does not exist any previous work that specifically proposed a framework to analyze data utility and information loss in privacy-preserving graph-formatted data. Therefore, this is the first framework intended to specifically evaluate data utility and privacy on graph-based data.

\subsection{Our contributions}

This paper proposes a common framework to evaluate data utility and information loss on privacy-preserving data publication processes. We will focus on simple, undirected and unlabelled graphs, since all other types of graphs can be easily converted to undirected graphs. Additionally, the literature on undirected graphs is far more extensive than the literature using other types of graphs.

Our main contributions are, but not limited to:

\begin{itemize}
    \item Our framework propounds some generic information loss measures (\ie application-independent) and some specific information loss measures (\ie application-dependent) on graph formatted data in order to provide a clear comparison between an original graph and several perturbed (\ie anonymous) graphs.
    
    \item We provide a wide range of well-known generic information loss measures (GIL), such as average degree, clustering coefficient or betweenness centrality, to evaluate to what extent the analysis of anonymized data differs from the original data in graph properties (\ie application-independent).
    
    \item We present specific information loss measures (SIL) to quantify data utility on application-dependent problems, such as community detection, information flow and detection of top influential users (experts).
    
    \item Additionally, we propose some metrics to evaluate the users' privacy considering an adversary that has degree-based knowledge and also an adversary that has information about the 1-neighborhood of each target vertex.
    
    \item We provide an implementation of our framework\footnote{Available at: \url{https://github.com/jcasasr/DUEF-GA}} licensed under the GNU General Public Licence.
    
    \item Finally, we present some example scenarios where our framework could be used by researchers and practitioners to choose the correct parametrization and/or algorithm.
\end{itemize}

\subsection{Notation}

Let $G=(V,E)$ be a simple, undirected and unlabelled graph, where $V$ is the set of vertices and $E$ the set of edges in $G$. We use $v_{i} \in V$ to denote vertex $i$ and $(v_{i}, v_{j}) \in E$ to indicate an edge connecting vertices $v_{i}$ and $v_{j}$. We define $n=\vert V \vert$ to refer to the number of vertices and $m=\vert E \vert$ to denote the number of edges. We use $G=(V,E)$ and $\widetilde{G}=(\widetilde{V},\widetilde{E})$ to indicate the original and the anonymized graphs, respectively.

\subsection{Roadmap}

This paper is organized as follows. We review the state of the art of anonymization in graphs in Section \ref{sec:preliminary-concepts} and graph assessment in Section \ref{sec:related-work}. Our framework is introduced in Section \ref{sec:framework}. Then, we discuss metrics related to generic information loss measures in Section \ref{sec:GIL} and a methodology to compare specific information loss measures in Section \ref{sec:SIL}. Metrics related to re-identification and risk assessment are detailed in Section \ref{sec:reidentification}. Finally, some example scenarios where our framework could be applied are presented in Section \ref{sec:examples}, while Section \ref{sec:conclusions} concludes the research and suggests directions for future work.

%% file: 2-concepts.tex
\section{Preliminary concepts}
\label{sec:preliminary-concepts}

In this section, we briefly describe the most important categories of anonymization methods in order to point out which ones could be used in our framework. According to \cite{Casas-RomaEtAl:2016:AIRE}, we categorize anonymization methods on graph-formatted data into three main categories:

\textbf{Graph modification approaches} anonymize a graph by modifying (adding and/or deleting) edges or vertices in a graph. These modifications could be done by adding and removing edges randomly, which is named randomization or random-based approach. There exists an extensive literature on graph randomization methods, such as \cite{HayEtAl:2007:TechRep,YingWu:2008:SDM,YingEtAl:2009:SNA-KDD,Casas-Roma:2014:MDAI,BonchiEtAl:2014:InfSci}. An alternative approach consists of edge addition and deletion to fulfill some desired constraints, \ie anonymization methods do not modify edges at random, they modify edges to meet some desired constraints. For instance, the $k$-anonymity methods \cite{Sweeney:2002:IJUFKS} modify graph structure in order to get the $k$-anonymity value of the graph. This model states that an attacker cannot distinguish among $k$ different records although he managed to find a group of quasi-identifiers. Consequently, the attacker cannot re-identify an individual with a probability greater than $\frac{1}{k}$. There are several works in literature on constrained anonymization based on edge modification, such as  \cite{LiuTerzi:2008:SIGMOD,ZhouPei:2008:ICDE,ZouEtAl:2009:VLDB,HayEtAl:2008:VLDB,Casas-RomaEtAl:2016:KAIS}. However, other works \cite{ChesterEtAl:2011:ADBIS,ChesterEtAl:2013:SNAM:a,MaEtAl:2015:Computing} permit modifications to the vertex and edge sets, rather than only to the edge set, and this offers some differences with respect to the utility of the released anonymous graph.

\textbf{Generalization approaches} cluster vertices and edges into super-vertices and super-edges to publish the aggregate information about the structural properties of the original graph \cite{HayEtAl:2008:VLDB}. Although users' privacy can be hidden properly, the graph may be shrunk considerably after anonymization. They do not enable local structure data analysis, so they are not a good approach to release data for specific and application-dependent tasks, such as detection of top influential users. Furthermore, it is very hard, if not impossible, to compare the structure of the original graphs to their generalized versions, since the number of vertices and edges could be drastically smaller than the original ones. The standard metrics used to analyze the graphs' structure, such as average distance or centrality measures, cannot be used to compare a graph and its generalized version.

Finally, \textbf{differentially private approaches} refer to methods which guarantee that individuals are protected under the definition of differential privacy \cite{Dwork:2006:ICALP}, which imposes a guarantee on the data release mechanism rather than on the data itself. Some works on differential privacy mechanism provide statistical information about the data, but they do not allow the anonimized graph to be released \cite{HayEtAl:2009:ICDM}. On the contrary, some other works use this concept to create and release an anonymized version of the original graph. For instance, the works by Sala et al. \cite{SalaEtAll:2011:IMC} and Brunet et al. \cite{BrunetEtAl:2016:IACR} release a perturbed version of the graph under the differential privacy constraints. Consequently, the former algorithms cannot be used in our framework, since they do not produce a perturbed version of the graph that could be compared to the original one. Contrariwise, the latter algorithms could be used in our framework to analyze the divergence between the original and the perturbed versions to analyze the information loss.

An extensive study about specific privacy-preserving methods and their particularities is beyond the scope of this work. However, some interesting surveys were made and can help to extend this brief summary, such as \cite{HayEtAl:2011:SIGMOD,Nagle:2013:MSNSI,Casas-RomaEtAl:2016:AIRE}.

In summary, we will focus on graph modification approaches and differential private methods that output a perturbed version of the graph, which both preserve local structures and keep the details of the data for subsequent graph-mining tasks. However this framework is not limited to these anonymization methods. Specifically, any privacy-preserving algorithm that produces and releases a perturbed version of the original graph could use this framework to study the data utility and the information loss.

%% file: 3-related-work.tex
\section{Related work}
\label{sec:related-work}

Several measures have been defined to quantify some properties of the graph structure \cite{BarabasiPosfai:2016:book}. Usually, these measures compare the values obtained by the original and the anonymized data in order to quantify the noise introduced by the anonymization process. However, the main problem is that there are several measures and, generally, different authors use different measures to compare their methods to the other state-of-the-art algorithms. We review the main measures used in literature in the following section.

Hay et al. \cite{HayEtAl:2007:TechRep} utilized two network-level properties (the degree distribution and the diameter) and three vertex-level measures (closeness centrality, betweenness centrality and path length distribution) from graph theory to quantify the information loss. Ying and Wu \cite{YingWu:2008:SDM} and Ying et al. \cite{YingEtAl:2009:SNA-KDD} used both real space and spectrum-based characteristics to study how the graph is affected by randomization methods. The authors focus on the harmonic mean of the shortest distance, modularity, transitivity, subgraph centrality, the largest eigenvalue of the adjacency matrix and the second smallest eigenvalue of the Laplacian matrix. Alternatively, Zou et al. \cite{ZouEtAl:2009:VLDB} defined a simple method based on the difference between the original and the anonymized edge set. Liu and Terzi \cite{LiuTerzi:2008:SIGMOD} used clustering coefficient and average path length for the same purpose. Finally, the degree, path length, clustering coefficient, network resilience and infectiousness were used by Hay et al. \cite{HayEtAl:2008:VLDB} to perform a similar analysis.

More recently, Casas-Roma \cite{Casas-Roma:2014:MDAI} carried out a data utility comparison among random methods by including several generic information loss measures: diameter, harmonic mean of the shortest distance, subgraph centrality, transitivity, core number sequence, betweenness and closeness centrality, the largest eigenvalue of the adjacency matrix and the second smallest eigenvalue of the Laplacian matrix. Wang et al. \cite{WangEtAl:2014:KAIS} used average path length, clustering coefficient, average betweenness and average closeness to compare their anonymous and original graphs. A very similar approach, using average path length, average clustering coefficient, average betweenness and percentage of changed edges, was followed by Macwan and Patel \cite{MacwanPatel:2018:CompJ}. An equivalent evaluation was performed by Wang and Zheng \cite{WangZheng:2015:ICDE} employing the degree, closeness, betweenness and PageRank centrality to evaluate information loss. Lindner et al. \cite{LindnerEtAl:2015:ASONAM} utilized diameter, normalized mutual information (NMI), vertex degree, betweenness centrality, PageRank and average local clustering coefficient to quantify data utility after sparsification to protect users' privacy in social networks. Bonchi et al. \cite{BonchiEtAl:2014:InfSci} performed a similar data utility analysis, based on four generic information loss measures (clustering coefficient, average distance, diameter and effective diameter). Additionally, the authors include an analysis of specific information loss measures related to epidemic threshold and clustering similarity.

As we have reviewed, the range of options to evaluate anonymous graphs is wide and it is complicated to compare among different methods and algorithms. A fair comparison among them is still an open problem that cannot be easily solved. Furthermore, just a few works included some task-specific information loss measures, such as community detection or information flow, which are important graph-mining tasks that will probably be performed on anonymous datasets.

%% file: 4-framework.tex
\section{Framework}
\label{sec:framework}

As we have previously stated, it might not be possible to compare data utility and information loss among relevant works in literature, since they are evaluated using different methodologies and metrics. Our objective is to provide a common framework to quantify information loss on graph perturbation processes. 

In this Section we provide an overall description of our framework, which includes two types of information loss measures and re-identification and risk assessment measures: 

\begin{itemize}
    \item We use \textbf{generic information loss measures (GIL)}, such as average distance or diameter, to evaluate to what extent the analysis of anonymized data differs from the original data. Each measure focuses on a particular property of the data. It is important to emphasize that these generic information loss measures only evaluate structural and spectral changes between original and anonymized data, and as such, they are general or application-independent. Details of the GIL measures are provided in Section \ref{sec:GIL}.
    
    \item We argue that the analysis of application-dependent and task-specific quality measures have to be considered to improve graph-mining tasks on anonymous datasets. Thus, we provide a set of \textbf{specific information loss measures (SIL)} designed to quantify perturbation on real-world specific tasks, such as community detection (\ie clustering), which we claim that should be considered to evaluate the goodness of the anonymization process. Section \ref{sec:SIL} presents the details of SIL measures considered in this work.
    
    \item Regarding the \textbf{re-identification and risk assessment measures}, Section \ref{sec:reidentification} points out some metrics to empirically evaluate the impact of external information on the adversary's ability to re-identify individuals.
\end{itemize}

As we have commented in Section \ref{sec:preliminary-concepts}, our framework focuses on any privacy-preserving algorithm that produces and releases a perturbed version of the original graph, including but not limited to graph modification and differential privacy methods. 

However, the released anonymous graphs must share some properties that we use to evaluate the data utility in our evaluation framework. Specifically, the vertex set must remain the same on perturbed graphs since some information loss metrics are computed for each vertex, \ie $V = \widetilde{V}$ and $n = \widetilde{n}$. Vertex modification methods could be used in our framework due to the fact that vertex addition could be represented as a change of the vertex's node degree from 0 to a positive number, and vertex removal changes the degree to 0. Nevertheless, our framework could not be used on generalization methods. On the contrary, the edge set changes due to the anonymization process, \ie $E \neq \widetilde{E}$, while the number of edges is usually modified, \ie $m \neq \widetilde{m}$. The number of edges on perturbed graphs could be greater than the original one, \ie $m < \widetilde{m}$, or smaller, \ie $m > \widetilde{m}$. Mainly, it depends on the algorithm's edge modification technique, which could be based on edge addition, deletion or a combination.

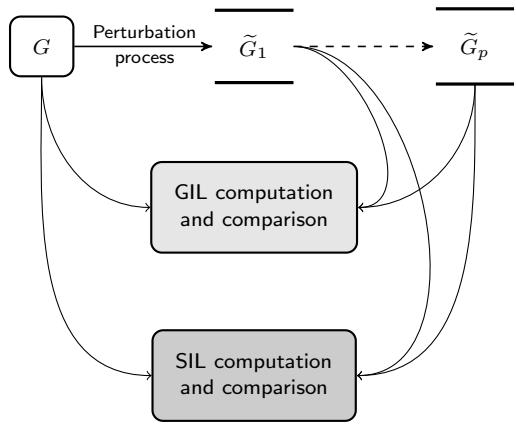
\begin{figure}
\centering
\begin{tikzpicture}[
		font=\sffamily\footnotesize,
  		every matrix/.style={ampersand replacement=\&,column sep=1cm, row sep=1cm},
  		source/.style={draw,thick,rounded corners,fill=white,inner sep=.3cm},
  		process/.style={draw,thick,circle,fill=black!20},
  		ga/.style={source,fill=black!10},
  		ca/.style={source,fill=black!20},
  		datastore/.style={draw,very thick,shape=datastore,inner sep=.3cm},
  		datastorega/.style={draw,very thick,shape=datastore,inner sep=.3cm,fill=black!10},
  		dots/.style={gray,scale=2},
  		to/.style={->,>=stealth',shorten >=1pt,semithick,font=\sffamily\scriptsize},
  		every node/.style={align=center}
	]

	\matrix
	{
    	\node[source] (dataori) {$G$}; \& \node[datastore] (data1) {$\widetilde{G}_1$}; \& \node[datastore] (data25) {$\widetilde{G}_p$}; \\
    		\& \node[ga] (graphassess) {GIL computation \\ and comparison}; \& \\
    		\& \node[ca] (clustassess) {SIL computation \\ and comparison}; \& \\
	};
  
	\draw[to] (dataori) -- node[midway,above] {Perturbation} node[midway,below] {process} (data1);
	\draw[to, dashed] (data1) -- (data25);
	
	\draw[->] (dataori) to[out=270,in=180] (graphassess);
	\draw[->] (data1) to[out=0,in=0] (graphassess);
	\draw[->] (data25) to[out=270,in=0] (graphassess);
	
	\draw[->] (dataori) to[out=270,in=180] (clustassess);
	\draw[->] (data1) to[out=0,in=0] (clustassess);
	\draw[->] (data25) to[out=270,in=0] (clustassess);
\end{tikzpicture}
	\caption{Experimental framework. The original dataset $G$ is perturbed to produce a sequence of anonymized graphs, \ie $\widetilde{G}_1, \ldots, \widetilde{G}_p$, using some anonymization method. Next, we compare the original and perturbed data using GIL measures in order to quantify the noise introduced on the data. Then, we do the same with real graph-mining processes and task-specific measures.}
	\label{fig:setup-1}
\end{figure}

Our experimental framework is shown in Figure \ref{fig:setup-1}. First, we apply perturbation to graph datasets using any graph modification approach. Each dataset is perturbed to produce a sequence of anonymized graphs, \ie $\widetilde{G}_1, \ldots, \widetilde{G}_p$. Then, we evaluate the original and the perturbed data using GIL measures for quantifying graph structure (details in Section \ref{sec:GIL}). Next, we apply real graph-mining tasks both on the original and the perturbed data and we use task-specific measures (Section \ref{sec:SIL}) to evaluate the results and quantify the divergence among them.

%% file: 5-GIL.tex
\section{Generic Information Loss Measures}
\label{sec:GIL}

In our framework, we use several generic measures to evaluate some key graph properties, which are relevant according to \cite{Casas-RomaEtAl:2014:KAIS}. They evaluate the graph structure, so they are general or, in other words, application-independent. These generic measures are used to quantify the noise introduced in the perturbed data by the anonymization process. The information loss is defined by the discrepancy between the results obtained from the original and the anonymized data. 

We use graph measures based on structural and spectral properties. Specifically, we consider three types of graph-related metrics: network-level, vertex-level and spectral metrics. In the rest of this section we review the selected measures. The number of vertices is not considered to assess anonymization process, since anonymization methods analyzed in this work keep these values constant. Other widely used metrics, such as diameter, are dismissed due to the fact that they proved to be worthless for data utility in perturbed graphs \cite{Casas-RomaEtAl:2014:KAIS}.

\subsection{Network-level metrics}

Network-level metrics are related to the whole graph, giving one score value for the entire graph. We consider the following ones:

\textbf{Average distance} ($\overline{dist}$) \cite{Casas-Roma:2014:MDAI,BonchiEtAl:2014:InfSci} is defined as the average of the distances between each pair of vertices in the graph. It measures the minimum average number of edges between any pair of vertices. Formally, it is defined as: 

\begin{equation}
\overline{dist}(G) = \frac{\sum_{i,j} d_{ij}} {\binom{n}{2}}
\end{equation}

\noindent where $d_{ij}$ is the length of the shortest geodesic path from $v_i$ to $v_j$, meaning the number of edges along the path.

\textbf{Clustering coefficient} \cite{LiuTerzi:2008:SIGMOD,HayEtAl:2008:VLDB,GirvanNewman:2002:PNAS,ChakrabartiFaloutsos:2006} (C) is a measure widely used in literature. The clustering coefficient of a graph is the average:

\begin{equation}
C(G) = \frac{1}{n} \sum_{i=1}^{n} C(v_{i}) 
\end{equation}

\noindent where $C(v_{i})$ is the clustering coefficient for vertex $v_{i}$. The clustering of each vertex is the fraction of possible triangles that exist. For each vertex the clustering coefficient is defined as:

\begin{equation}
C(v_{i}) = \frac{2T(v_{i})}{deg(v_{i})(deg(v_{i})-1)}
\end{equation}

\noindent where $T(v_{i})$ is the number of triangles surrounding vertex $v_{i}$ and $deg(v_{i})$ is the degree of $v_{i}$.

\textbf{Transitivity} \cite{YingWu:2008:SDM,YingEtAl:2009:SNA-KDD,ChakrabartiFaloutsos:2006} (T) is the fraction of all possible triangles present in the graph. Possible triangles are identified by the number of triads (two edges with a shared vertex), as we can see in Equation \ref{eq:T}.

\begin{equation}
\label{eq:T}
T(G) = \frac{3 \times (number \, of \, triangles)}{(number \, of \, triads)}
\end{equation}

As we have previously stated, these metrics compute a score value for the whole graph. Thus, we calculate the error on these metrics by applying Equation \ref{eq:network-diff}, where $m$ stands for the particular metric.

\begin{equation}
\epsilon_m(G,\widetilde{G}) = \vert m(G) - m(\widetilde{G}) \vert
\label{eq:network-diff}
\end{equation}

Finally, another widely used measure is \textbf{edge intersection} \cite{ZouEtAl:2009:VLDB,LiuTerzi:2008:SIGMOD} (EI). It is defined as the percentage of original edges which are also in the anonymized graph. Formally:

\begin{equation}
EI(G,\widetilde{G}) = \frac{\vert E \cap \widetilde{E} \vert}{max(\vert E \vert,\vert \widetilde{E} \vert)}
\end{equation}

We do not use Equation \ref{eq:network-diff} since this metric is intrinsically defined to quantify divergence between two different graphs. 

\subsection{Vertex-level metrics}

Vertex-level metrics compute a score value for each vertex in the graph. In order to facilitate comparison among different graphs, it would be convenient to have a single score value for a graph. The undermentioned metrics are proposed:

\textbf{Betweenness centrality} \cite{HayEtAl:2007:TechRep} (BC) is a centrality measure, which calculates the fraction of the number of the shortest paths that go through each vertex. This measure indicates the centrality of a vertex based on the flow among other vertices in the graph. The betweenness centrality of vertex $v_{i}$ is defined as:

\begin{equation}
BC(v_{i}) = \frac{1}{n^{2}} \sum_{s,t} \frac{g^{i}_{st}}{g_{st}}
\end{equation}

\noindent where $g^{i}_{st}$ is the number of geodesic paths from $v_s$ to $v_t$ that pass through $v_{i}$, and $g_{st}$ is the total number of geodesic paths from $v_s$ to $v_t$.

The second centrality measure is \textbf{closeness centrality} \cite{HayEtAl:2007:TechRep} (CC), which is described as the inverse of the average distance to all accessible vertices. Thus, a larger value indicates less centrality, while a smaller value indicates higher centrality. Formally, the closeness centrality of a vertex $v_{i}$ is defined as:

\begin{equation}
CC(v_{i}) = \frac{n}{\sum_{j} d_{ij}}
\end{equation}

And the last centrality measure is \textbf{degree centrality} \cite{HayEtAl:2007:TechRep} (DC), which evaluates the centrality of each vertex through its degree. The degree centrality of a vertex $v_{i}$ is depicted in Equation \ref{eq:DC}.

\begin{equation}
\label{eq:DC}
DC(v_{i}) = \frac{deg(v_{i})}{m}
\end{equation}

The last three centrality measures described above evaluate the centrality of each vertex of the graph from different perspectives. These measures give us a value of centrality for each vertex. Therefore, we compute the vector of differences for each vertex between the original and the anonymized graph and the root mean square (\textit{RMS}) to obtain a single value for the whole graph, as showed in Equation \ref{eq:vertex-diff}.

\begin{equation}
\epsilon_m(G,\widetilde{G}) = \sqrt{\frac{1}{n} \sum_{i=1}^{n}{(g_i - \widetilde{g}_i)^{2}}}
\label{eq:vertex-diff}
\end{equation}

\noindent where $m$ is the specific vertex-level metric, $g_{i}$ is the value of the centrality measure for the vertex $v_i$ of $G$, and $\widetilde{g}_{i}$ is the value of the centrality measure for the vertex $v_i$ of $\widetilde{G}$. In our experiments we use Equation \ref{eq:vertex-diff} to compute a value representing the error induced in the whole graph by the anonymization process in the centrality measures.

\subsection{Spectral metrics}

We focus on the \textbf{largest eigenvalue of the adjacency matrix A} ($\lambda_{1}$) \cite{YingWu:2008:SDM} where $\lambda_{i}$ are the eigenvalues of A and $\lambda_{1} \geq \lambda_{2} \geq \ldots \geq \lambda_{n}$. The eigenvalues of A encode information about the cycles of a graph as well as its diameter. The spectral decomposition of A is:

\begin{equation}
A = \sum_{i} \lambda_{i} e_{i} e_{i}^{T}
\end{equation}

\noindent where $e_{i}$ is the eigenvector corresponding to $\lambda_{i}$ eigenvalue.

Only one score value is provided for each graph, since we only consider the first eigenvalue ($\lambda_{1}$). Thus, we compute the error on this metric using the same method as network-level metrics, \ie Equation \ref{eq:network-diff}.

\subsection{Complexity}

Time and space complexity are important issues when dealing with large or very large networks. Some of the aforementioned metrics, such as degree centrality, takes $\mathcal{O}(m)$ time. However, measures related to triads, like clustering coefficient or transitivity, need $\mathcal(n^2)$ running time. Calculating the average distance, betweenness centrality and closeness centrality implies calculating the shortest paths between all pairs of vertices, which could be very expensive to compute for large graphs. This takes $\mathcal{O}(n^3)$ time with the Floyd-Warshall algorithm. On sparse graphs, Johnson's algorithm may be more efficient, which takes $\mathcal{O}(n^2 log(n) + n m)$ time, and $\mathcal{O}(nm)$ using Brandes' algorithm on undirected and unweighted graphs \cite{Brandes:2001}. Finally, the spectral decomposition of a matrix takes $\mathcal{O}(n^3)$ time, which makes $\lambda_{1}$ infeasible for large or very large networks.

%% file: 6-SIL.tex
\section{Specific Information Loss Measures}
\label{sec:SIL}

Variations in the generic graph properties are a good way to assess the information loss but they have their limitations because they are just a proxy to the changes in data utility we actually want to measure. For instance, the average distance or the diameter could remain constant while the topology of the graph completely changes at the vertex level. What we are truly interested in is, given a data mining task at hand, quantify the disparity in the results between performing the task on the original graph and on the anonymized one.

\subsection{Clustering or community detection}
\label{sec:SIL-clustering}

We chose clustering because it is an active field of research, which provides interesting and useful information in community detection, for instance. Therefore, the extracted clusters/communities of vertices are the data utility we want to preserve. Like generic information loss measures, we compute the information loss by quantifying the divergence between the original and the perturbed data. This measure is task-specific and application-dependent, but we claim it is important to test the perturbed data in real community detection processes.

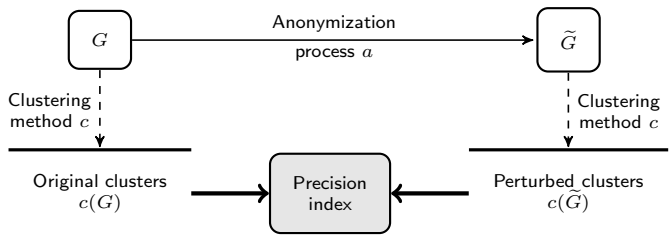
\begin{figure}
\centering
\begin{tikzpicture}[
		font=\sffamily\scriptsize,
  		every matrix/.style={ampersand replacement=\&,column sep=1cm, row sep=1cm},
  		source/.style={draw,thick,rounded corners,fill=white,inner sep=.3cm},
  		process/.style={draw,thick,circle,fill=black!20},
  		ga/.style={source,fill=black!10},
  		ca/.style={source,fill=black!20},
  		datastore/.style={draw,very thick,shape=datastore,inner sep=.3cm},
  		datastorega/.style={draw,very thick,shape=datastore,inner sep=.3cm,fill=black!10},
  		dots/.style={gray,scale=2},
  		to/.style={->, >=stealth', shorten >=1pt, semithick},
  		every node/.style={align=center}
	]

	\matrix
	{
    	\node[source] (G) {$G$}; \& \& \node[source] (Gb) {$\widetilde{G}$}; \\
    	\node[datastore] (cG) {Original clusters \\ $c(G)$}; \& \node[ga] (precision) {Precision \\ index}; \& \node[datastore] (cGb) {Perturbed clusters \\ $c(\widetilde{G})$}; \\
	};
	
	\draw[to] (G) -- node[midway,above] {Anonymization} node[midway,below] {process $a$} (Gb);
	\draw[to, dashed] (G) -- node[midway,left] {Clustering\\ method $c$} (cG);
	\draw[to, dashed] (Gb) -- node[midway,right] {Clustering\\ method $c$} (cGb);
	\draw[->, ultra thick, black] (cG) -- (precision);
	\draw[<-, ultra thick, black] (precision) -- (cGb);
\end{tikzpicture}
	\caption{Framework for evaluating the clustering-specific information loss measure.}
	\label{fig:setup-SIL-clustering}
\end{figure}

We considered the following approach to evaluate the clustering assignment made by a given clustering method $c$ using a particular graph perturbation method $a$: 

\begin{enumerate}
    \item Apply $a$ to the original data $G$ and obtain $\widetilde{G} = a(G)$.
    \item Apply $c$ to $G$ and $\widetilde{G}$ to obtain the cluster assignments $c(G)$ and $c(\widetilde{G})$.
    \item Compare $c(G)$ to $c(\widetilde{G})$ to evaluate the divergence between them.
\end{enumerate}

The approach is illustrated in Figure \ref{fig:setup-SIL-clustering}. In terms of information loss, it is clear that the more similar $c(\widetilde{G})$ is to $c(G)$, the less information loss. Thus, clustering-specific information loss metrics should measure the divergence between both cluster assignments $c(G)$ and $c(\widetilde{G})$. Ideally, if the anonymization step was lossless in terms of data utility, we should have the same number of clusters with the same elements in each cluster. When the clusters do not match, we use the \textit{precision index} \cite{CaiEtAl:2010:ICMLC} to quantify the divergence. Assuming we know the true communities of a graph, the precision index can be directly used to evaluate the similarity between two cluster assignments. Given a graph of $n$ vertices and $q$ true communities, we assigned to vertices the same labels $l_{tc}(\cdot)$ as the community they belong to. In our case, the true communities are the ones assigned to the original dataset (\ie $c(G)$) since we want to obtain communities as close as the ones we would get on non-anonymized data. Assuming the perturbed graph has been divided into clusters (\ie $c(\widetilde{G})$), then for every cluster, we examine all the vertices within it and assign to them as predicted label $l_{pc}(\cdot)$ the most frequent true label in that cluster (basically the mode). Then, the precision index can be defined as follows:

\begin{equation}
precision\_index(G,\widetilde{G}) = \frac{1}{n} \sum_{v \in G} \mathds{1}_{l_{tc}(v) = l_{pc}(v)}
\label{eq:precision}
\end{equation}

\noindent where $\mathds{1}$ is the indicator function such that $\mathds{1}_{x=y}$ equals 1 if $x=y$ and 0 otherwise. Note that the precision index is a value in the range [0,1], which takes value 0 when there is no overlap between the sets and value 1 when the overlap between the sets is complete. To be consistent with the notion of error for the generic graph properties, we report 1 - $precision\_index$ in the results tables so that the lower, the better.

Regarding the clustering methods $c$, we propose 4 graph clustering algorithms to evaluate the edge modification techniques. All of them are unsupervised algorithms based on different concepts and developed for different applications and scopes. An extended revision and comparison of them can be found in Lancichinetti and Fortunato \cite{LancichinettiFortunato:2009:PhysRevE} and Zhang et al. \cite{ZhangEtAl:2013:KAIS}. The selected clustering algorithms are:

\begin{itemize}
\item \textbf{Infomap} by Rosvall and Bergstrom \cite{RosvallBergstrom:2008:PNAS} use the problem of optimally compressing the information on the structure of the graph to find the best cluster structure. This is achieved by compressing the information of a dynamic process taking place on the graph, namely a random walk. The optimal compression is achieved by optimizing a quality function, which is the Minimum Description Length of the random walk. Such optimization can be carried out rather quickly with a combination of greedy search and simulated annealing. It runs in time $\mathcal{O}(m)$.

\item Fast greedy modularity optimization (\textbf{Fastgreedy}) by Clauset, Newman and Moore \cite{ClausetEtAl:2004:PhysRevE} is a hierarchical agglomeration algorithm for detecting community structure based on modularity optimization. Starting from a set of isolated vertices, the edges of the original graph are iteratively added to produce the largest possible increase of the modularity at each step. Its running time on a sparse graph is $\mathcal{O}(n \log^{2} n)$.

\item \textbf{Multilevel} by Blondel et al. \cite{BlondelEtAl:2008:JSTAT} is a multi-step technique based on a local optimization of Newman-Girvan modularity in the neighborhood of each vertex. After a partition is identified in this way, communities are replaced by super-vertices, yielding a smaller weighted graph. The procedure is then iterated, until modularity does not increase any further. The computational complexity is essentially linear in the number of edges of the graph, \ie $\mathcal{O}(m)$.

\item \textbf{Walktrap} \cite{PonsLatapy:2005:ISCIS} by Pons and Latapy tries to find densely connected subgraphs, also called communities in a graph via random walks. The idea is that short random walks tend to stay in the same community. They proposed a measure of similarities between vertices based on random walks to capture the community structure in a graph. It runs in time $\mathcal{O}(mn^{2})$ and space $\mathcal{O}(n^{2})$ in the worst case.
\end{itemize}

Walktrap and Infomap are based on the random walk concept, while Fastgreedy is based on hierarchical edge betweenness and Multilevel is based on modularity concept. Even though some algorithms permit overlapping among different clusters, we did not allow it in our experiments by setting the corresponding parameter to zero, mainly for ease of evaluation.

\subsection{Remaining Ratio of Top Influential Users}
\label{sec:SIL-RRTI}


We also chose a score metric related to the remaining ratio of top influential users (RRTI). This metric is quite important in several areas of business, such as marketing. Viral marketing seeks to maximize the spread of a campaign through an online social network, often targeting influential users with high centrality. Therefore, preserving the most important users is relevant in graph-mining tasks. In this case, the top influential users are the data utility we want to preserve on anonymous datasets.

We propose to test the remaining ratio of top influential users to show how the published graph preserves this utility. Let us denote the set of top $x$ percent influential users in $G$ as $\Theta_x(G)$, and the corresponding one in $\widetilde{G}$ as $\Theta_x(\widetilde{G})$. We define the remaining ratio of the top influential users as:

\begin{equation}
    RRTI(G,\widetilde{G}) = \frac{\vert \Theta_x(G) \cap \Theta_x(\widetilde{G}) \vert}{\vert \Theta_x(G) \vert}
    \label{eq:rrti}
\end{equation}

Note that the RRTI is a value in the range [0,1], which takes value 0 when there is no overlap between the sets and value 1 when the overlap between the sets is complete. The larger RRTI is, the better the published graph preserves the information in the original graph.

In our framework we use the PageRank algorithm \cite{PageEtAl:1998:WWW} to compute the users' influential values. Additionally, we use the set of top 20 percent influential users in our experiments \cite{YuanEtAl:2013:TKDE} though it can be modified according to the graph size.

There are simple and fast random walk-based distributed algorithms for computing PageRank of vertices in a graph. The authors in \cite{Sarma:2015:TCS} present a simple algorithm that takes $\mathcal{O}(log~n / \epsilon)$ rounds with high probability on any graph (directed or undirected), where $\epsilon$  is the reset probability ($1-\epsilon$ is also called as \textit{damping factor}) used in the PageRank computation. They also present a faster algorithm that takes $\mathcal{O}(\sqrt{log~n} / \epsilon)$ rounds in undirected graphs. Both of the above algorithms are scalable to large and very large graphs.

\subsection{Information Flow}
\label{sec:SIL-flow}

Finally, another very important graph-mining task is related to information flow and spreading phenomena, \ie how information (or a disease) spreads in a graph. In this context, several works have been published using epidemic modelling, such as susceptible-infected (SI) model, susceptible-infected-susceptible (SIS) model or susceptible-infected-recovered (SIR) model. However, these models do not consider the underlying graph structure, and are usually based on the degree distribution of the graph. Furthermore, they are designed to be used on directed and weighted graphs. Consequently, they are not easy to adapt to simple, unlabelled and undirected graphs.

We propose an approach to evaluate the information flow based on the distance between each pair of vertices in the graph. Specifically, we focus on the maximum distance from each vertex to reach all other vertices in the graph. We claim that this measure is important to define how information flows (or a disease spreads) in a graph. The process is as follows:

\begin{enumerate}
    \item Compute the distance from each vertex $v_i$ to all other vertices and keep the maximum one, as described by:
    
    \begin{equation}
        \delta(v_i) = dist_{max}(v_i, v_j) : v_i \neq v_j
    \end{equation}
    
    \item Create vector $\Delta(G) = (\delta(v_1), \ldots, \delta(v_n)) : v_i \in G$ and, equivalently, $\Delta(\widetilde{G})$.
    
    \item Compute the divergence between two vectors to quantify the divergence between vectors related to the farthest reachable vertex (FRV) by using the following equation:.
    
    \begin{equation}
        FRV(G,\widetilde{G}) = \frac{\sum \vert \Delta(G) - \Delta(\widetilde{G}) \vert}{n}
        \label{eq:flow_loss}
    \end{equation}
\end{enumerate}

Note that many algorithms can be used to compute the distances between vertices. In our case, breadth-first search (BFS) is used since edges are unweighted. The time complexity of BFS can be expressed as $\mathcal{O}(n + m)$ and the space complexity is $\mathcal{O}(n)$.

In terms of information loss, it is clear that the more similar $\Delta(\widetilde{G})$ is to $\Delta(G)$, the less information loss. Ideally, if the anonymization step was lossless in terms of data utility, we would have the same vectors. However, we need a metric to quantify the distance between vectors. For this purpose, we propose to use Equation \ref{eq:flow_loss}.

Note that this metric takes value 0 when vectors are identical. On the contrary, values $\approx n$ indicate that each vertex has increased or decreased by $n$ the distance to the furthest vertex in the graph (on average). For instance, a value $\approx 1$ shows that, on average, information starting on a randomly chosen vertex will need $\pm 1$ steps to reach all other vertices in the graph.

%% file: 7-reidentification.tex
\section{Re-identification and risk assessment}
\label{sec:reidentification}


The privacy level is usually defined through the anonymization method. For instance, the $k$ value in $k$-anonymity methods specifies the probability to re-identify a user in the anonymous dataset. However, we present some metrics to evaluate the privacy level in a common way in order to ease the comparison among different methods and techniques.

Particularly, the metrics we propose are designed to empirically evaluate the impact of external information on the adversary's ability to re-identify individuals. For each data set, we consider each vertex in turn as a target. We propose two scenarios, considering an adversary with different external information: first, we consider that the adversary has degree-based knowledge and, second, we assume the adversary has information about the 1-neighborhood of each target vertex. An extended revision of privacy metrics can be found at \cite{WagnerEckhoff:2018:CS,YangEtAl:2012:CODASPY}.

Similar to the data utility measures, these ones can be used on any anonymized graph obtained by graph modification approaches or differential privacy methods. Nevertheless, they are very relevant when dealing with random-based graphs, since they can point out interesting information under these models.

\subsection{Degree-based attacks}
\label{sec:degree-based}

In this section we consider an attacker with degree-based knowledge. Two different scenarios are provided to evaluate the privacy.

\subsubsection{Vertices that changed their degree}

Firstly, we propound a simple metric that computes the number of vertices which changed their degree during the perturbation or anonymization process. It is clear that the higher the proportion of vertices that changed their degree, the harder the re-identification process will be. 

The global percentage of the anonymization could report information about the general perturbation of the graph, though it does not provide information about how the perturbation affects each vertex of the graph, \ie the perturbation could be concentrated in a few vertices of the graph, leaving the other vertices without alteration. Thus, this metric is useful to analyze how the local structure of the graph changed during the anonymization process.

\subsubsection{Candidate set size}

Secondly, we propose a metric based on the concept of anonymity set size \cite{WagnerEckhoff:2018:CS}. The anonymity set for an individual $v_i$ is the set of users that the adversary cannot distinguish from $v_i$, which can be seen as the size of the crowd into which the target $v_i$ can be. 

We assume the adversary computes a vertex refinement query on each vertex, and then computes the corresponding candidate set for each vertex. We report the distribution of candidate set sizes across the population of vertices to characterize how many vertices are protected and how many are identifiable. According to Hay et al. \cite{HayEtAl:2008:VLDB}, the candidate set of a target vertex $v_i$ includes all vertices $v_j \in \widetilde{G}$ such that $v_j$ is a candidate in some possible world. Assuming that the anonymization method and the number $w$ of fake edges are public, the adversary must consider the set of possible worlds implied by $\widetilde{G}$ and $w$. Informally, the set of possible worlds consists of all graphs that could result in $\widetilde{G}$ under $w$ perturbations.

Using $E^c$ to refer to all edges not present in $E$, the set of possible worlds of $\widetilde{G}$ under $w$ random edge perturbations, denoted by $\mathcal{W}^{w}(\widetilde{E})$, corresponds to:

\begin{equation}
\mathcal{W}^{w}(\widetilde{E}) = \binom{\vert E \vert}{w} \binom{\vert E^c \vert}{w}
\end{equation}

\noindent where $\vert E \vert = m$ is the number of edges, $\vert E^c \vert = \frac{n (n-1)}{2} - m$ is the number of edges of the graph complement.

We compute the candidate set of the target vertex $v_i$ based on vertex refinement queries of level 1 ($\mathcal{H}_1$) (see Hay et al. \cite{HayEtAl:2008:VLDB} for further details) as shown in Equation \ref{eq:candH1}.

\begin{equation}
Cand_{\mathcal{H}_1}(v_i) = \{v_j : deg^{-}(v_j) \leq deg(v_i) \leq deg^{+}(v_j) \}
\label{eq:candH1}
\end{equation}

\begin{equation}
deg^{-}(v_j) = \Big\Vert deg(v_j) \Big(1 - \frac{w}{\vert E \vert}\Big) \Big\Vert
\label{eq:degmin}
\end{equation}

\begin{equation}
deg^{+}(v_j) = \Big\Vert deg(v_j) + (n-1-deg(v_j)) \Big( \frac{w}{\vert E^c \vert} \Big) \Big\Vert
\label{eq:degmax}
\end{equation}

\noindent where $deg^{-}(v_j)$ is the minimum expected degree of $v_j$ and $deg^{+}(v_j)$ is the maximum expected degree after $w$ edge deletion/addition process, and described by the following equations. Note that $deg^{-}(v_j) = deg(v_j)$ when the anonymization is based on edge addition, and $deg^{+}(v_j) = deg(v_j)$ when the anonymization is based on edge deletion.

Following the work by Hay et al. \cite{HayEtAl:2008:VLDB}, we propose five groups based on the number of vertices whose equivalent candidate set size falls into each of them: [1] (note that these users can be directly re-identified), [2,4] (groups of 2 to 4 users), [5,10], [11,20] and [21, $\infty$] (well-protected users).

\subsection{1-Neighborhood-based attacks}
\label{sec:1-neighborhood-based}

Finally, we consider an adversary with 1-neighborhood-based knowledge \cite{ZhouPei:2008:ICDE}, \ie an adversary who knows the friends' set of some target vertices. Similar to the previous measures, we computed the proportions of vertices that change their set of neighbors at distance one during the perturbation process. 

As we have previously stated, it is important to obtain information about how the perturbation affects each vertex of the graph. Specifically, how the perturbation alters the sub-graphs at distance 1 of each target vertex.

%% file: 8-examples.tex
\section{Application examples}
\label{sec:examples}

In this section we briefly present some hypothetical results obtained by our experimental framework and anonymous graphs obtained from randomization and $k$-degree anonymous algorithms.

\subsection{Tested Graphs}
\label{sec:datasets}

We use three different real graphs in our experiments. Table \ref{table:comparison-graphs-properties} shows a summary of their main features. They are the following ones:

\begin{table}[!t]
	\caption{Graph properties. For each dataset we present the number of vertices ($n$), number of edges ($m$), average degree ($\overline{deg}$), average distance ($\overline{dist}$), diameter ($D$) and the exponent of the fitted power-law distribution ($\gamma$).}
	\label{table:comparison-graphs-properties}
	\centering{}
	\renewcommand\tabcolsep{5pt}
	\def\arraystretch{1.2}
	{\footnotesize
	\begin{tabular}{ l | r | r | r | r | r | r }
		\textbf{Dataset} & $~n~$ & $~m~$ & $~\overline{deg}~$ & $~\overline{dist}~$ & $~D~$ & $~\gamma~$ \\
		\hline
		\hline
		Infectious	& 410 & 2,765 & 13.487 & 3.630 & 9 & 6.423 \\
		\hline
		URV email 	& 1,133 & 5,451 & 9.622 & 3.606 & 8 & 6.775 \\
		\hline
		Hamsterster & 1,858	& 12,534 & 13.491 & 3.452 & 14 & 4.319 \\
		\hline
	\end{tabular}
	}
\end{table}

\begin{itemize}
	\item Infectious \cite{konect:sociopatterns} describes the face-to-face behavior of people during the exhibition ``Infectious: stay away'' in 2009 at the Science Gallery in Dublin. Vertices represent exhibition visitors and edges represent face-to-face contacts that were active for at least 20 seconds.
	
	\item URV email \cite{GuimeraEtAl:2003:PhysRevE} is the email communication network at the University Rovira i Virgili in Tarragona (Spain). Vertices are users and each edge represents that at least one email has been sent.
	
	\item Hamsterster friendships \cite{konect:2017:petster-friendships-hamster} contain friendships between users of the website hamsterster.com.
\end{itemize}

\subsection{Scenario I: Analyzing an anonymous graph}

The framework expects two input graphs, $G$ and $\widetilde{G}$, and it returns a score value for each GIL and SIL metrics. We choose Infectious as a dataset for this first experiment. We have compared the original Infectious graph to a 10\% randomly perturbed version ($\widetilde{G}_{10}$) generated by using edge addition \cite{Casas-RomaEtAl:2016:AIRE}. We run 10 independent executions of the perturbation algorithm to provide statistical significance of the results.

\begin{table*}[!t]
    \centering{}
    \caption{Data utility sample results for scenario I. The values of the perturbed dataset (second row) present the mean value ($\mu$) and the 95\% confidence interval (CI) of 10 independent executions.}
    \def\arraystretch{1.3}
    \setlength{\tabcolsep}{4pt}
    {\footnotesize
    	\begin{tabular}{ c c || r | r | r | r || r | r | r | r || r || r}
    		\textbf{Graph} & & $\overline{dist}$ & $EI$ & $BC$ & $\lambda_1$ & $Infomap$ & $Multilevel$ & $Fastgreedy$ & $Walktrap$ & $RRTI$ & $FRV$ \\
            \hline
            \hline
    		$G$ &  & 3.630 & 1.000 & 0.000 & 23.382 &
    		    1.000 & 1.000 & 1.000 & 1.000 & 1.000 & 0.000 \\
    		\hline
    		$\widetilde{G}_{10}$ & $\mu$ & 2.933 & 0.909 & 0.016 & 24.686 &
    		    0.933 & 0.918 & 0.863 & 0.875 & 0.878 & 1.946 \\
    		& CI & $\pm$ 0.0097 & $\pm$ 0.0001 & $\pm$ 0.0002 & $\pm$ 0.0338 &
    		    $\pm$ 0.0264 & $\pm$ 0.0386 & $\pm$ 0.0680 & $\pm$ 0.0295 & $\pm$ 0.0123 & $\pm$ 0.0611 \\
    		\hline
    	\end{tabular}
    }
    \label{table:s1}
\end{table*}

Sample results are depicted in Table \ref{table:s1}. We have reported the original score values (first row) and the values of the perturbed dataset (second row). We report the mean value ($\mu$) and the 95\% confidence interval (CI) for each metric on the perturbed (\ie anonymized) dataset. Only some GIL metrics results have been included due to space constraints. Some reported GIL scores show relevant information about the original values and the perturbed ones, such as average distance or $\lambda_1$. Other measures, like edge intersection or betweenness centrality, point out relative error to the original graph.

Regarding our clustering-specific information loss measures, we can see the precision index on our four clustering or community detection algorithms. We can see that Infomap seems to be more robust to the perturbation introduced during the anonymization process. It achieves not only the best mean score, but also the lowest confidence interval error. On the contrary, precision index on Fastgreedy algorithm performs the worst. Concerning the other specific information loss measures, our framework demonstrates that more than 87\% of the most important users remain in the perturbed graph, while the distance of the furthest vertex changes in $\pm 1.946$ steps.

Results of the candidate set size ($Cand_{\mathcal{H}_1}$) are depicted in Table \ref{table:s1_candH1}, where we can see that the number of users in high risk of re-identification decreases considerably in the perturbed graph ($\widetilde{G}_{10}$), while the number of well-protected users increases from 78 to 318.

\begin{table}[!t]
    \centering{}
    \caption{Candidate set size ($Cand_{\mathcal{H}_1}$) results for scenario I.}
    \def\arraystretch{1.3}
    \setlength{\tabcolsep}{4pt}
    {\footnotesize
    	\begin{tabular}{ c || r | r | r | r | r }
    		\textbf{Graph} & [1] & [2,4] & [5,10] & [11,20] & [21, $\infty$] \\
            \hline
            \hline
    		$G$ & 4 & 17 & 78 & 233 & 78 \\
    		\hline
    		$\widetilde{G}_{10}$ & 2 & 10 & 19 & 61 & 318 \\
    		\hline
    	\end{tabular}
    }
    \label{table:s1_candH1}
\end{table}

Finally, the number of vertices that changed their degree during the anonymization process is 301.3 with a confidence interval (95\% CI) of $\pm 6.0309$, which indicates that 75\% of vertices are hinder to re-identify using degree-based knowledge. The analysis of 1-neighborhood-based attacks presents similar results.

\subsection{Scenario II: Evaluating a graph sequence}

In this case, the framework expects a sequence of $p$ input graphs, \ie $\{G, \widetilde{G_1} \ldots, \widetilde{G}_{p-1}\}$, and it returns a score value for each metric and graph.

We chose URV email data to illustrate this scenario. The sequence of input graphs includes the original URV email dataset and some $k$-degree anonymous versions, where $k \in \{2, \ldots, 10\}$ (note that $k=1$ corresponds to the original dataset). Anonymization was performed by UMGA algorithm \cite{Casas-RomaEtAl:2016:KAIS}, which is based on the concept of $k$-degree anonymity. Furthermore, it is important to underline that this algorithm is deterministic and, therefore, it is not necessary to run independent executions of the algorithm. Consequently, it is not necessary to compute the statistical significance of the results.

When applying a $k$-anonymous algorithm, the privacy level is defined according to the quasi-identifiers and the parameter $k$. To provide a good anonymous dataset, one has to achieve a good trade-off between privacy and data utility. In order to do so, it is useful to analyze how data utility and information loss behave during the anonymization process. It can help to choose the correct k parameter related to not only privacy aspects, but also data utility in subsequent graph-mining processes.

\begin{figure*}[!t]
  \center
  \subfloat[Edge intersection]{
  	\label{fig:s2-EI}
  	\includegraphics[width=0.23\linewidth]{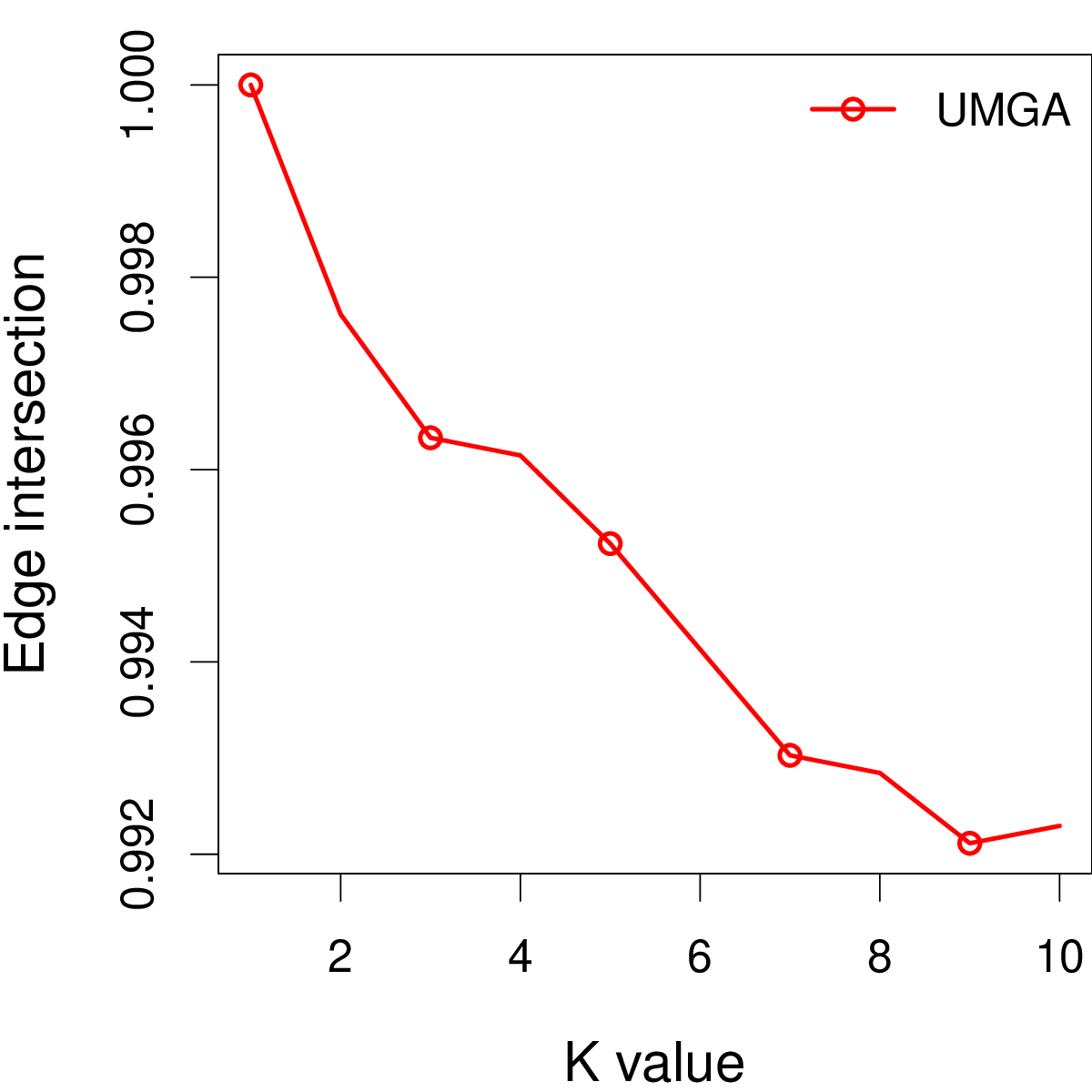}
  }
  ~
  \subfloat[Closeness centrality]{
  	\label{fig:s2-CC}
  	\includegraphics[width=0.23\linewidth]{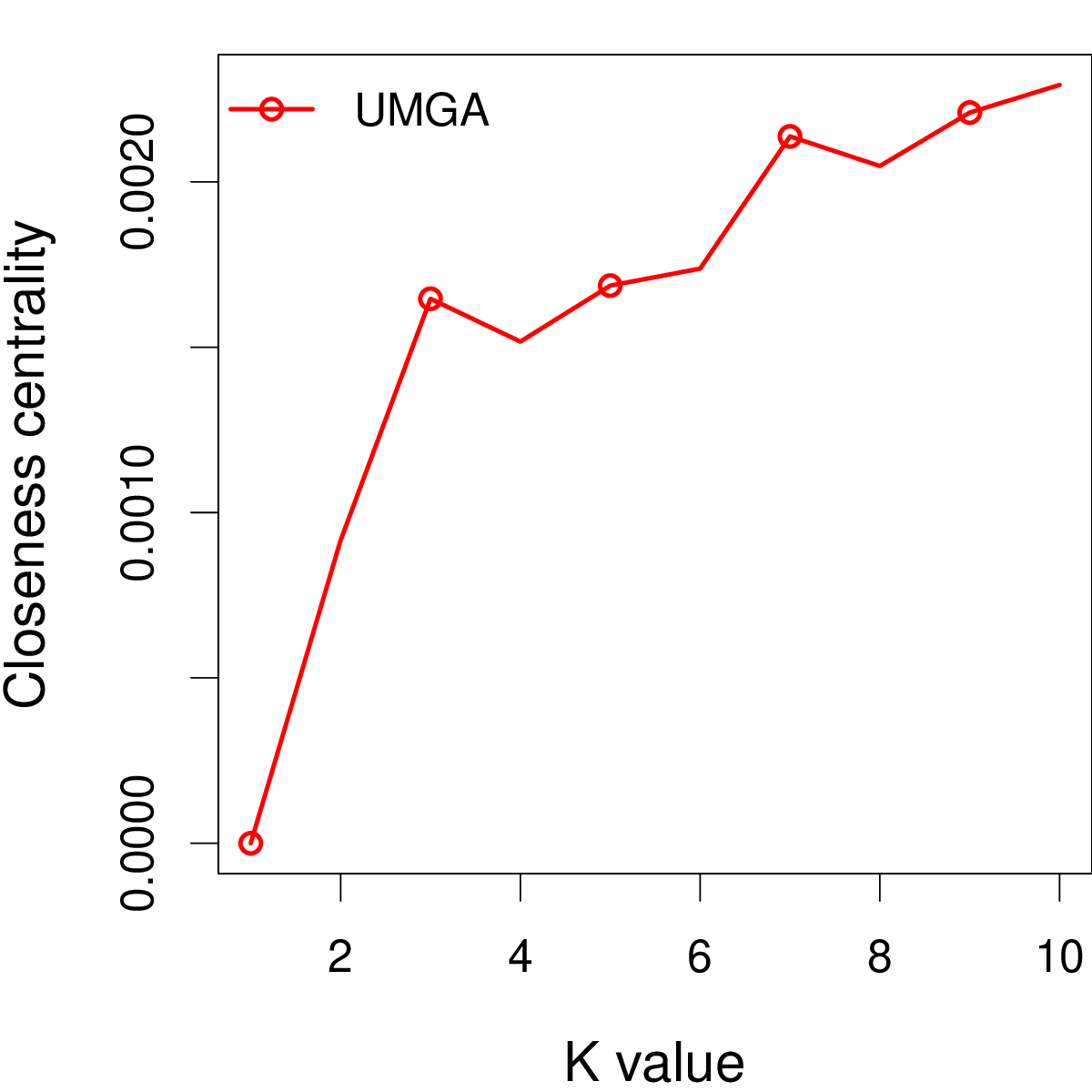}
  }
  ~
  \subfloat[Precision index (Multilevel)]{
  	\label{fig:s2-ML}
  	\includegraphics[width=0.23\linewidth]{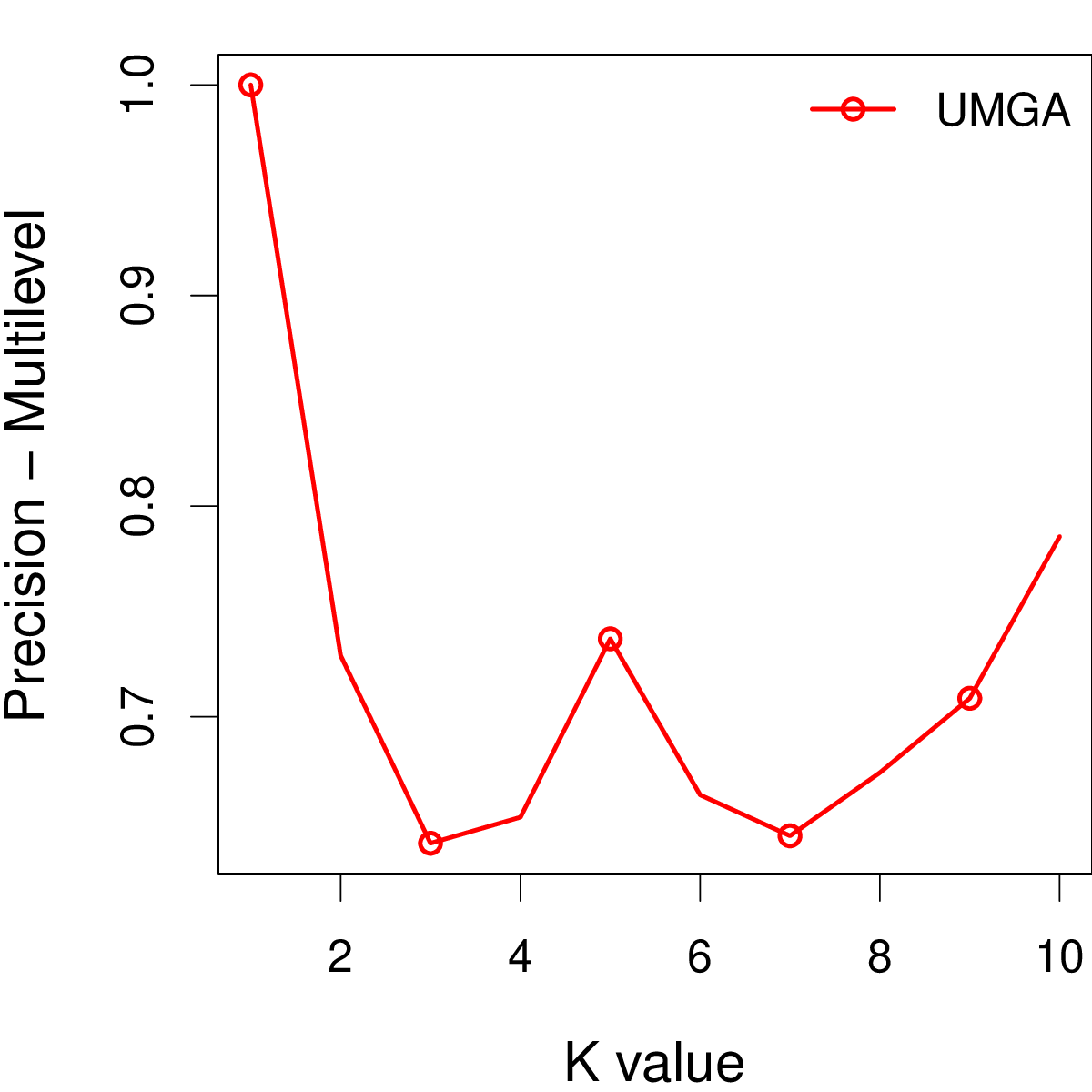}
  }
  ~
  \subfloat[RRTI]{
  	\label{fig:s2-RRTI}
  	\includegraphics[width=0.23\linewidth]{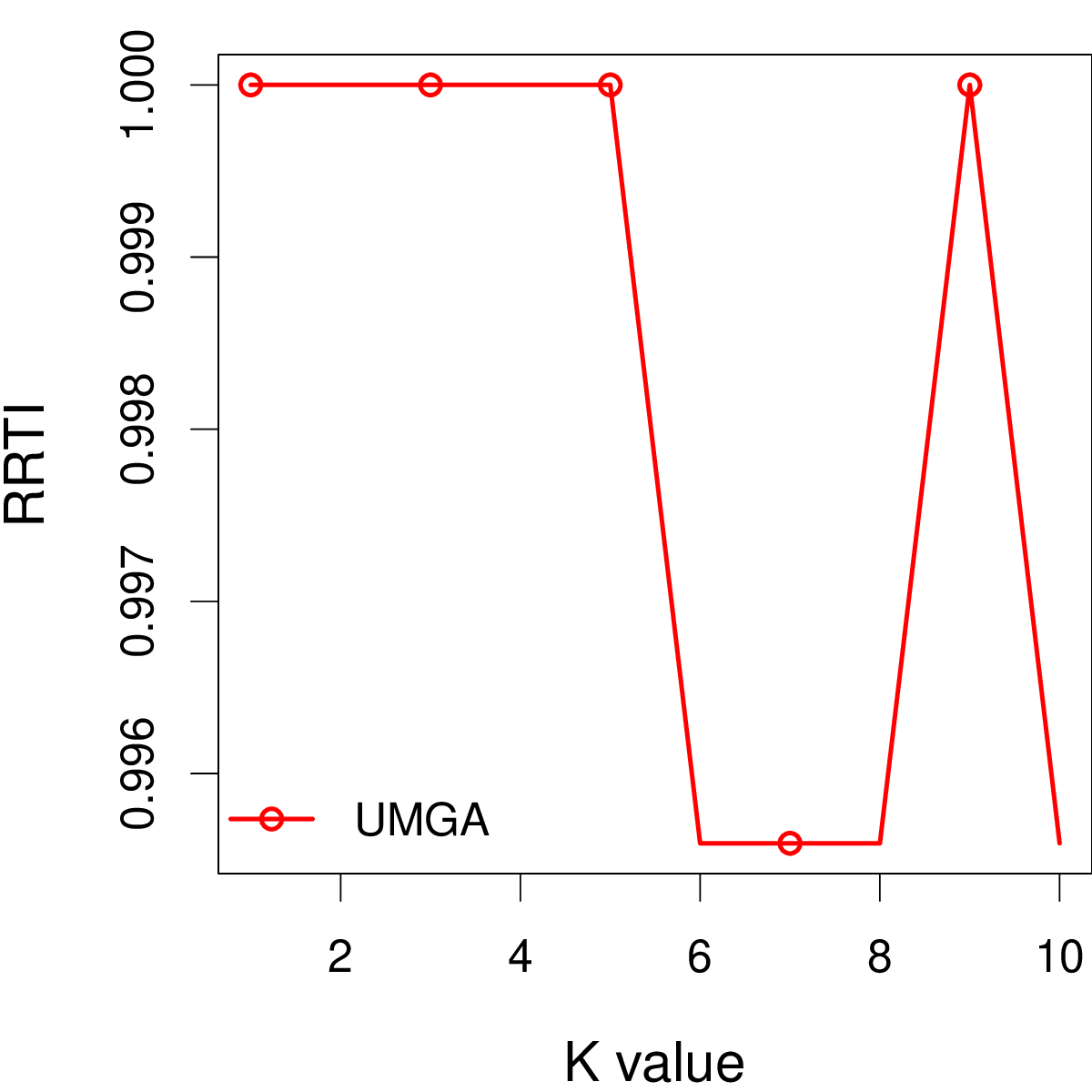}
  }
  \\
  \subfloat[Information flow]{
  	\label{fig:s2-FRV}
  	\includegraphics[width=0.23\linewidth]{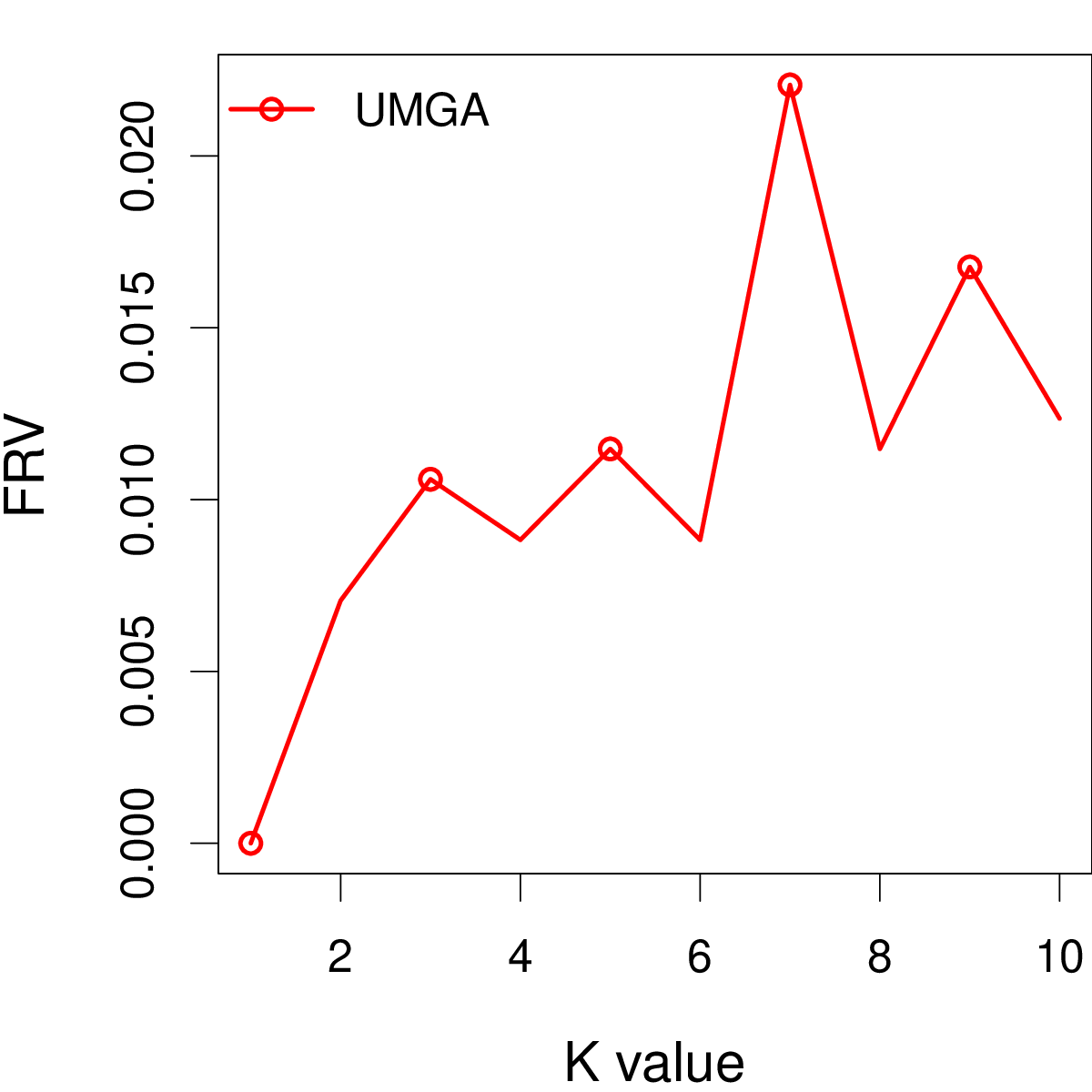}
  }
  ~
  \subfloat[$Cand_{\mathcal{H}_1}$]{
  	\label{fig:sample-case-2-H1}
  	\includegraphics[width=0.23\linewidth]{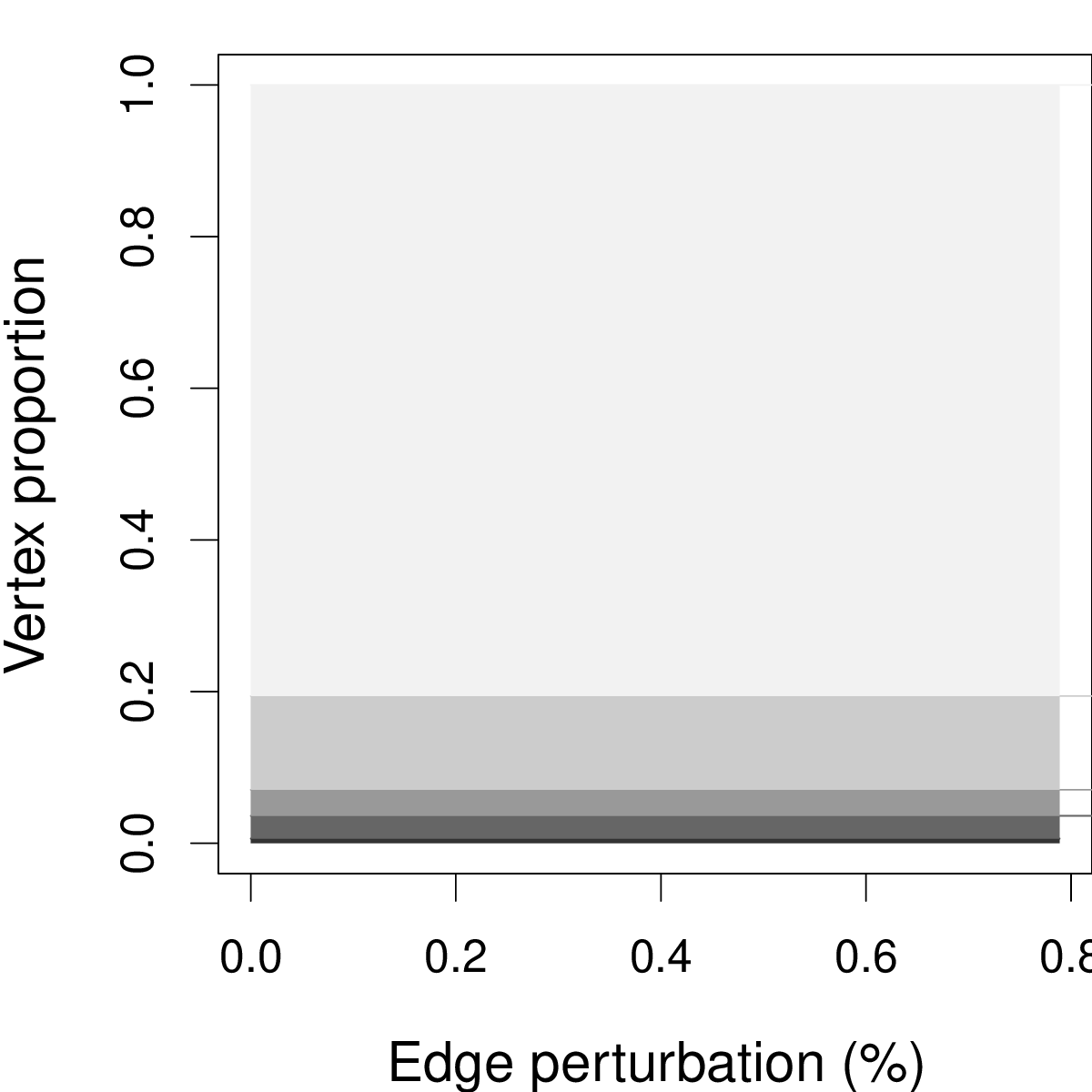}
  }
  ~
  \subfloat[Vertices that changed their degree]{
  	\label{fig:s2-DCV}
  	\includegraphics[width=0.23\linewidth]{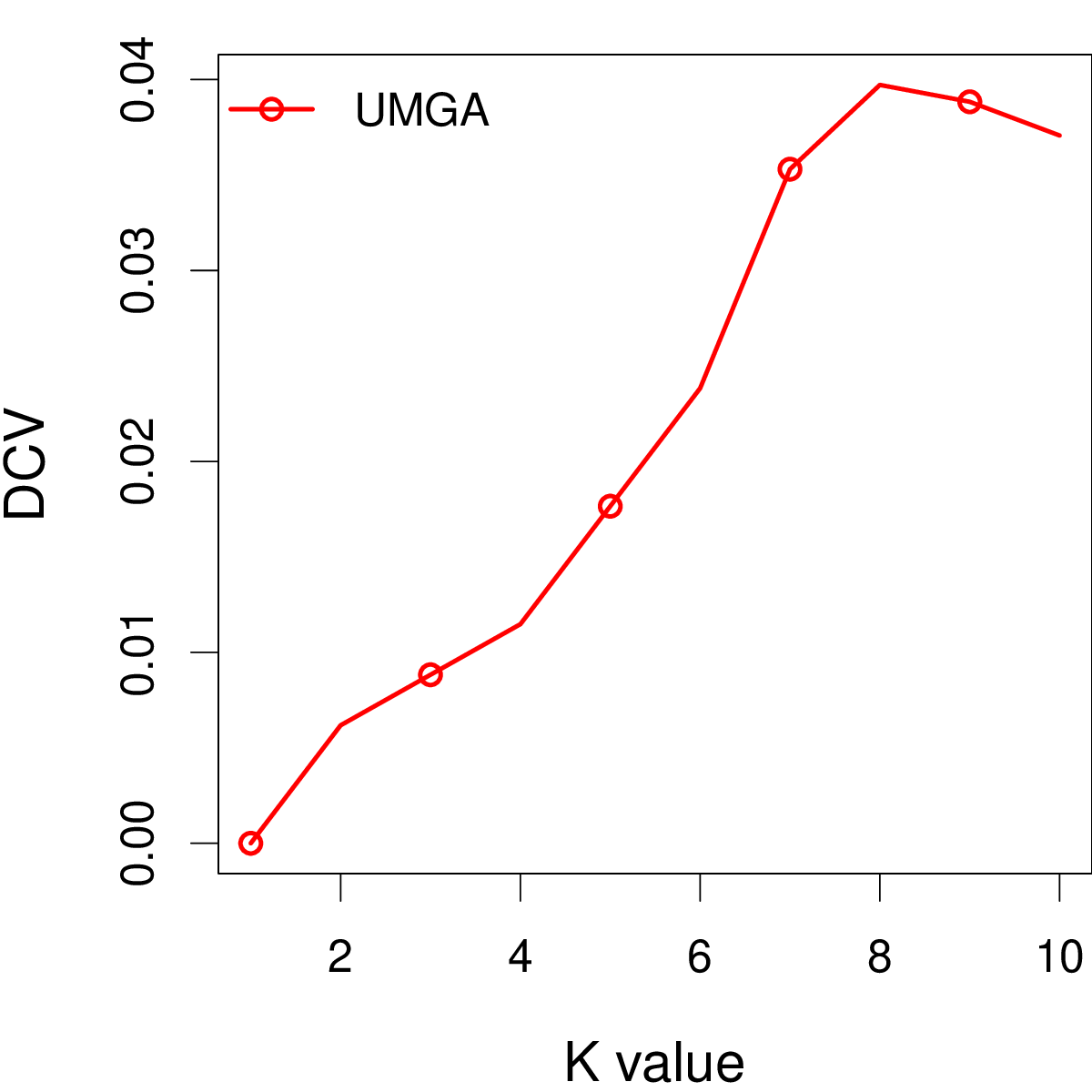}
  }
  ~
  \subfloat[1-Neighbourhood]{
  	\label{fig:s2-Neigh}
  	\includegraphics[width=0.23\linewidth]{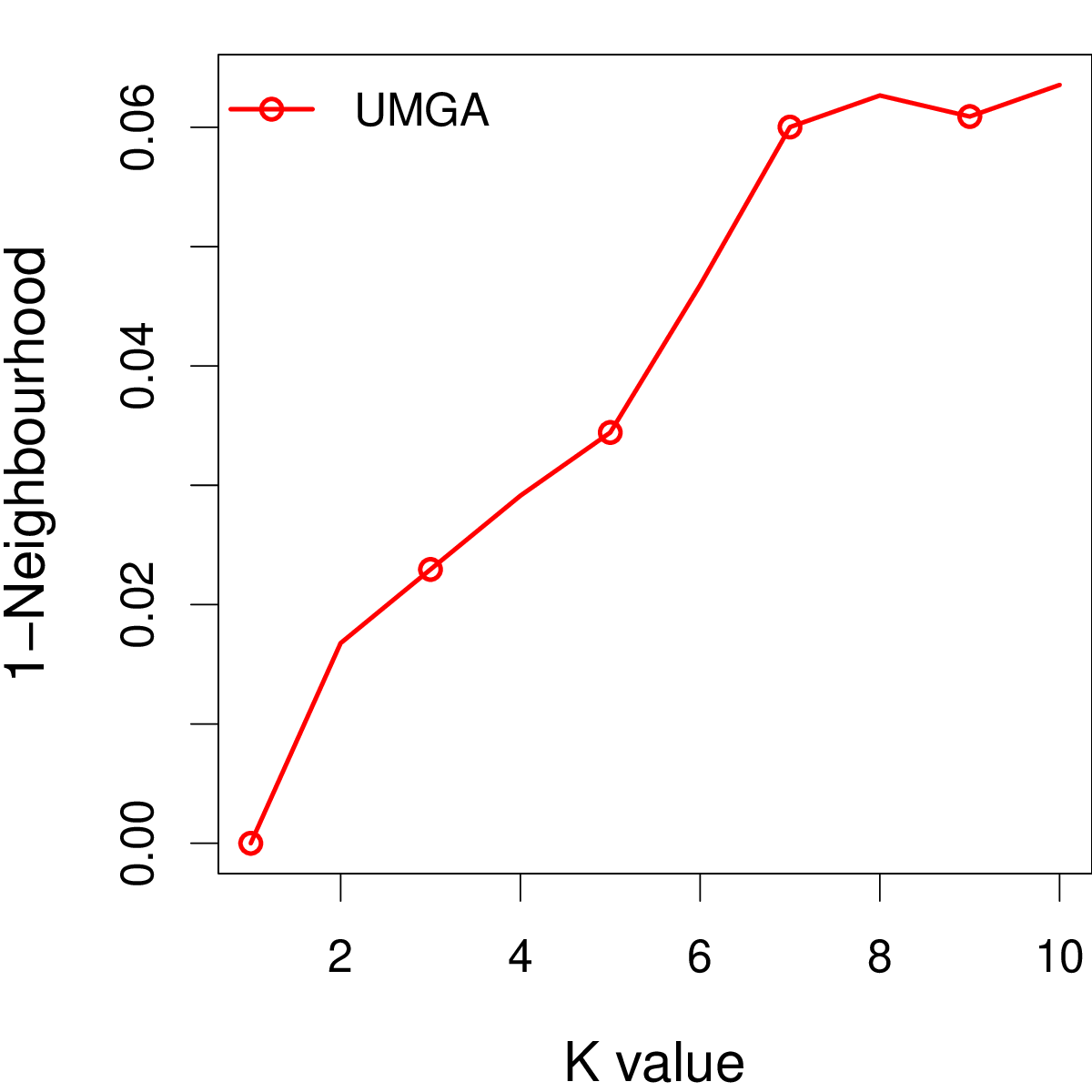}
  }
  \caption{Examples of our framework results for Scenario II. The horizontal axis presents the anonymization level ($k$-anonymity value), while vertical axis indicates the value of the original graph (leftmost point) and the evolution during anonymization processes.}
  \label{fig:s2}
\end{figure*}

The data utility and privacy results are depicted in Figure \ref{fig:s2}. Generally, the greater the privacy, the lower the data utility. Figure \ref{fig:s2-EI} points out that the number of fake edges is very small. As we can see, the UMGA algorithm achieves a privacy level of $k=10$ while keeping more than 99.2\% of the edges from the original graph, \ie altering 0.77\% of the edge set. Therefore, we achieve a reasonable level of privacy at a very low cost. Closeness centrality, depicted in Figure \ref{fig:s2-CC}, presents similar information loss for values of $k \geq 3$. The precision index using the Multilevel algorithm (Figure \ref{fig:s2-ML}) presents high information loss in all anonymization range. On the contrary, RRTI keeps all the most important users in the anonymous graphs (Figure \ref{fig:s2-RRTI}) and FRV indicates a rather small perturbation on information flow (Figure \ref{fig:s2-FRV}).

Regarding the re-identification and risk assessment analysis, we present the candidate set size results in Figure \ref{fig:sample-case-2-H1}. Perturbation varies along the horizontal axis and vertex proportion is represented on the vertical axis. The trend lines show the proportions of vertices whose equivalent candidate set size falls into each of the following groups: [1] (black), [2,4], [5,10], [11,20], [21, $\infty$] (light grey). We cannot appreciate differences in the groups, due to the fact that the UMGA algorithm perturbs less than 0.8\% of the total number of edges during the anonymization process. Finally, the number of vertices that changed their degree (Figure \ref{fig:s2-DCV}) and also the number of vertices that changed their 1-neighborhood (Figure \ref{fig:s2-Neigh}) point out that $k \geq 7$ achieve better results on both metrics.

In conclusion, using the UMGA algorithm and URV email dataset, and according to our experiments, we suggest a parametrization of $k=10$, since it provides a good anonymization level while keeping data utility close to their original values. We claim that our framework can help to choose the correct parameters to set up any privacy-preserving algorithm by comparing several information loss measures (both generic and specific) between original and anonymous datasets.

\subsection{Scenario III: Comparing multiple graphs sequences}

In this third scenario, the framework expects two or more sequences of $p$ input graphs, \ie $\{GS_1, \ldots, GS_q\} : q \ge 2$ where $GS_i$ stands for the $i$-th sequence of graphs and corresponds to $p$ input graphs $GS_i = \{G, \widetilde{G_1^w} \ldots, \widetilde{G}_{p-1}^w\}$, and it returns a score value for each metric and graph. Note that $\widetilde{G}_{j}^w$ represents the $j$-th perturbed version of graph $G$ using algorithm and parametrization $w$.

Researchers are interested in comparing the original graph to a set of anonymous graphs obtained from the original one (\ie $\widetilde{G}_1, \ldots, \widetilde{G}_p$), by applying different percentages of randomization or different $k$ values in $k$-anonymity-based algorithms. Not only may this help to understand the behavior of the datasets, but also to choose the best parameter according to privacy and data utility requirements. 

\begin{figure*}[!t]
  \center
  \subfloat[Average distance]{
  	\label{fig:s3-AD}
  	\includegraphics[width=0.23\linewidth]{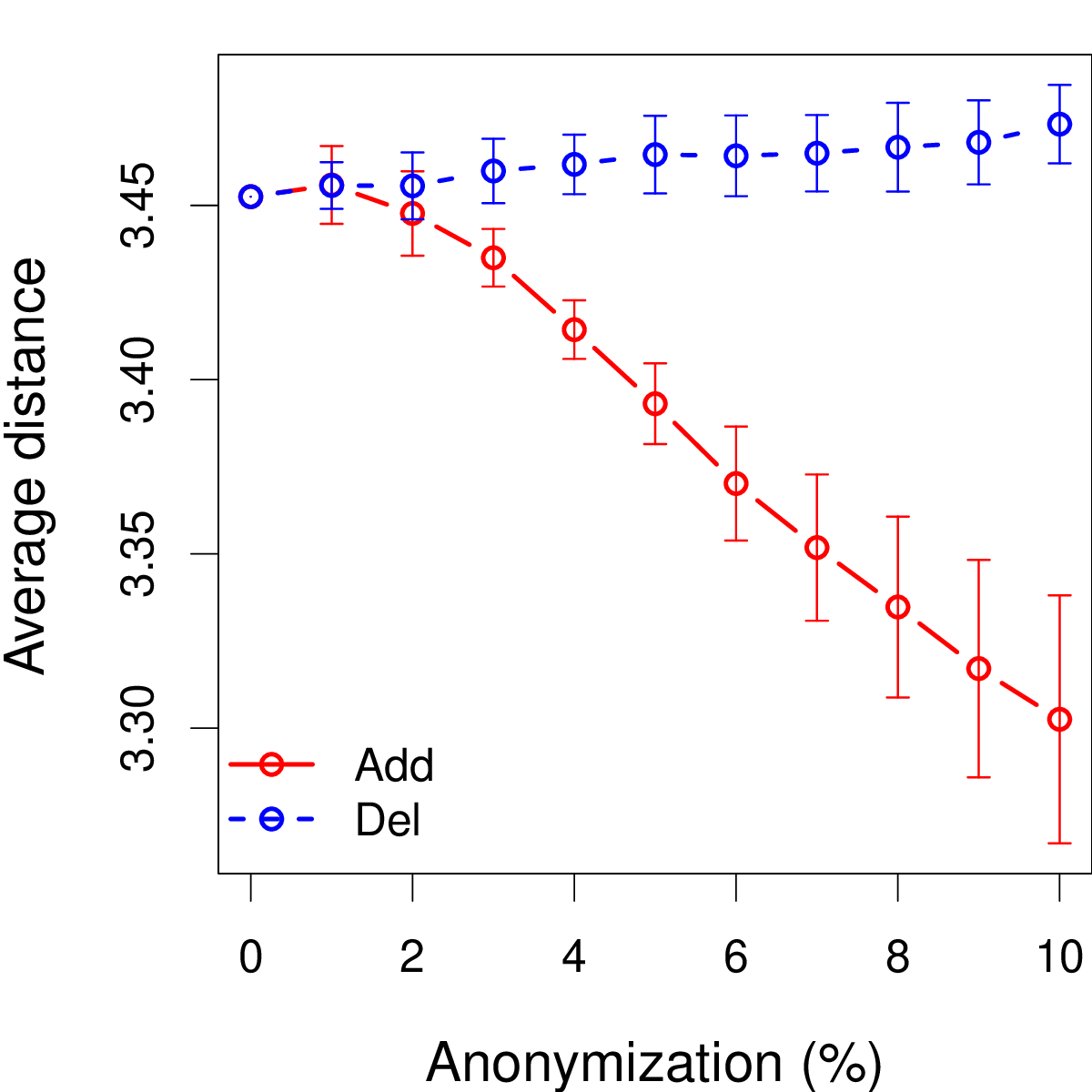}
  }
  ~
  \subfloat[Betweenness centrality]{
  	\label{fig:s3-BC}
  	\includegraphics[width=0.23\linewidth]{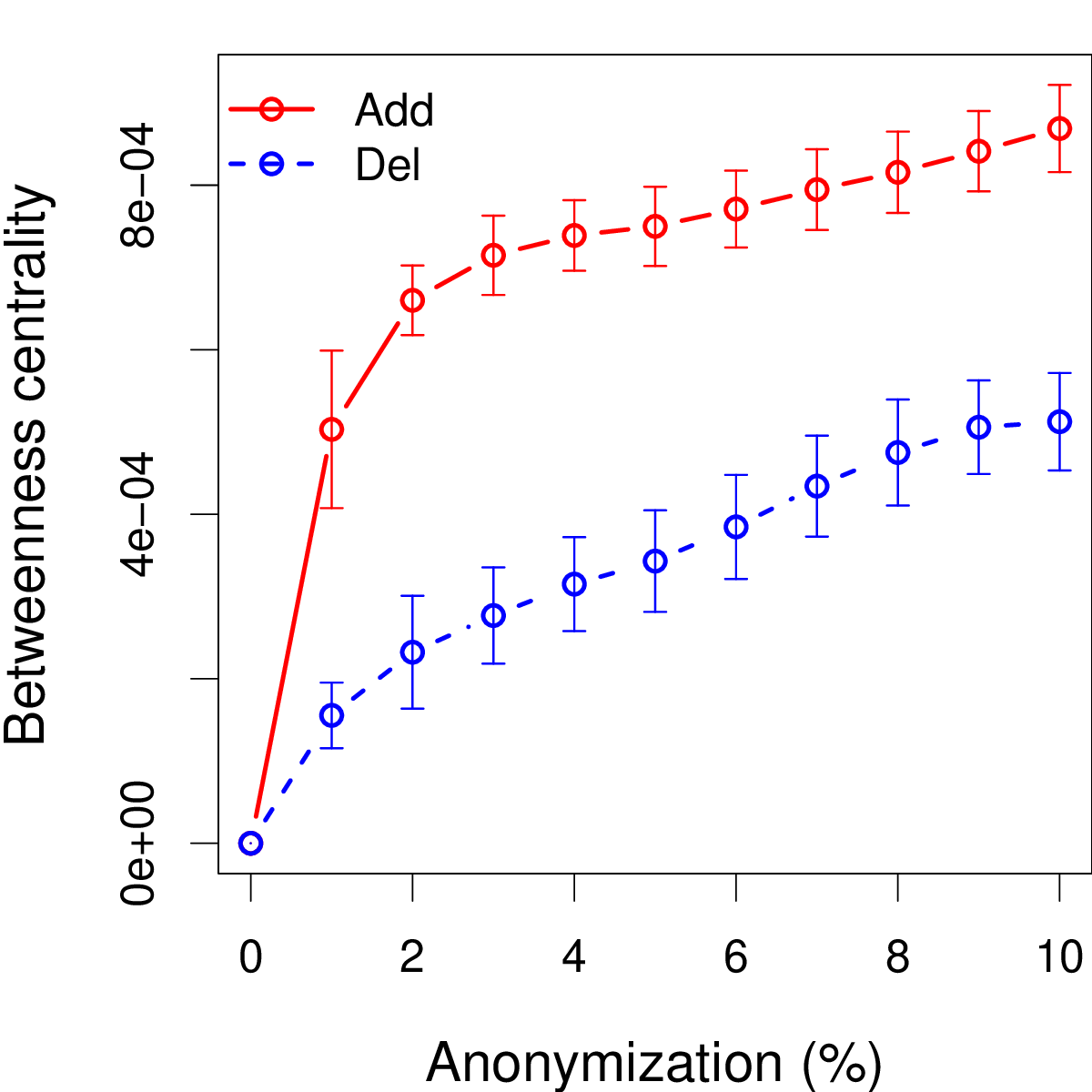}
  }
  ~
  \subfloat[Precision index (Walktrap)]{
  	\label{fig:s3-WT}
  	\includegraphics[width=0.23\linewidth]{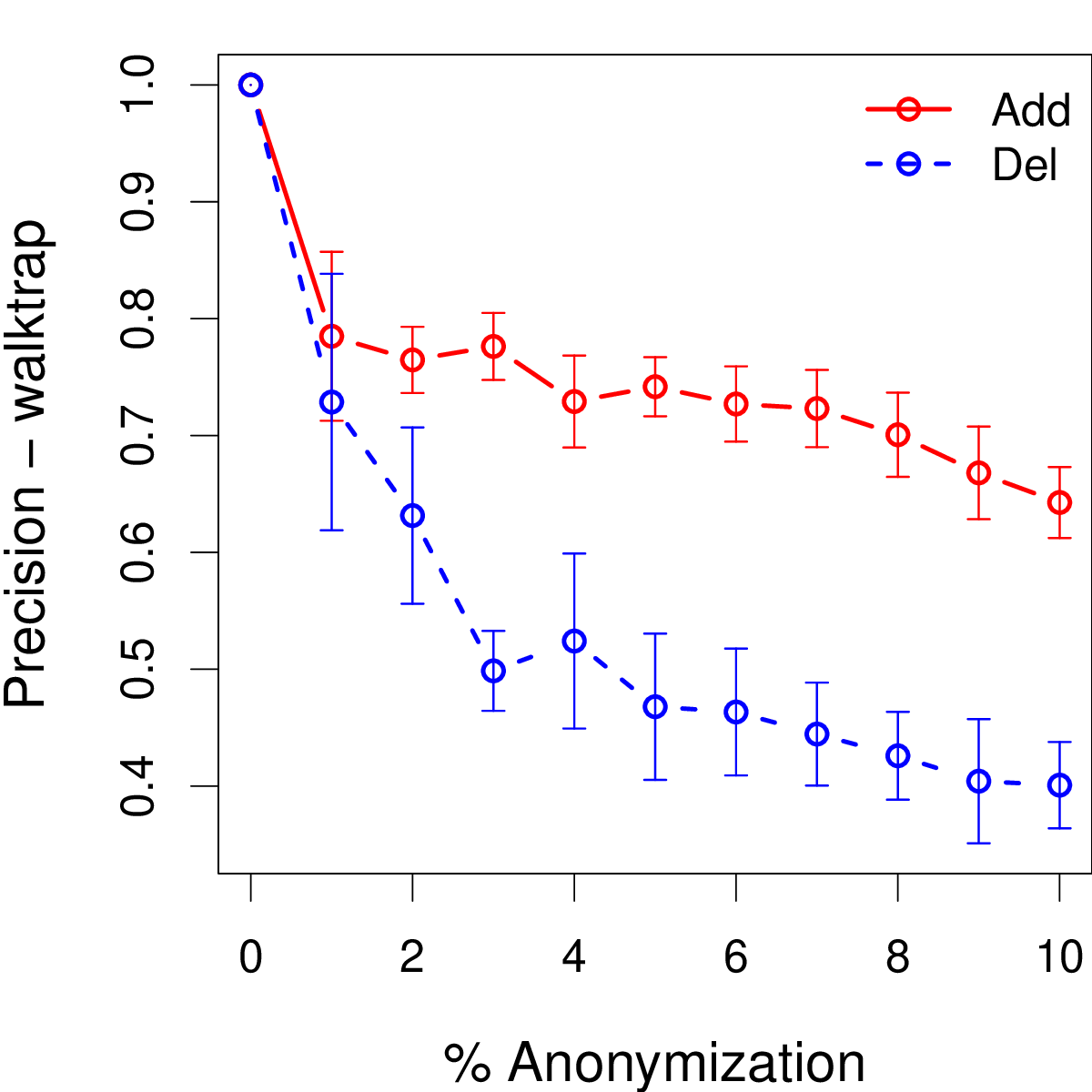}
  }
  ~
  \subfloat[RRTI]{
  	\label{fig:s3-RRTI}
  	\includegraphics[width=0.23\linewidth]{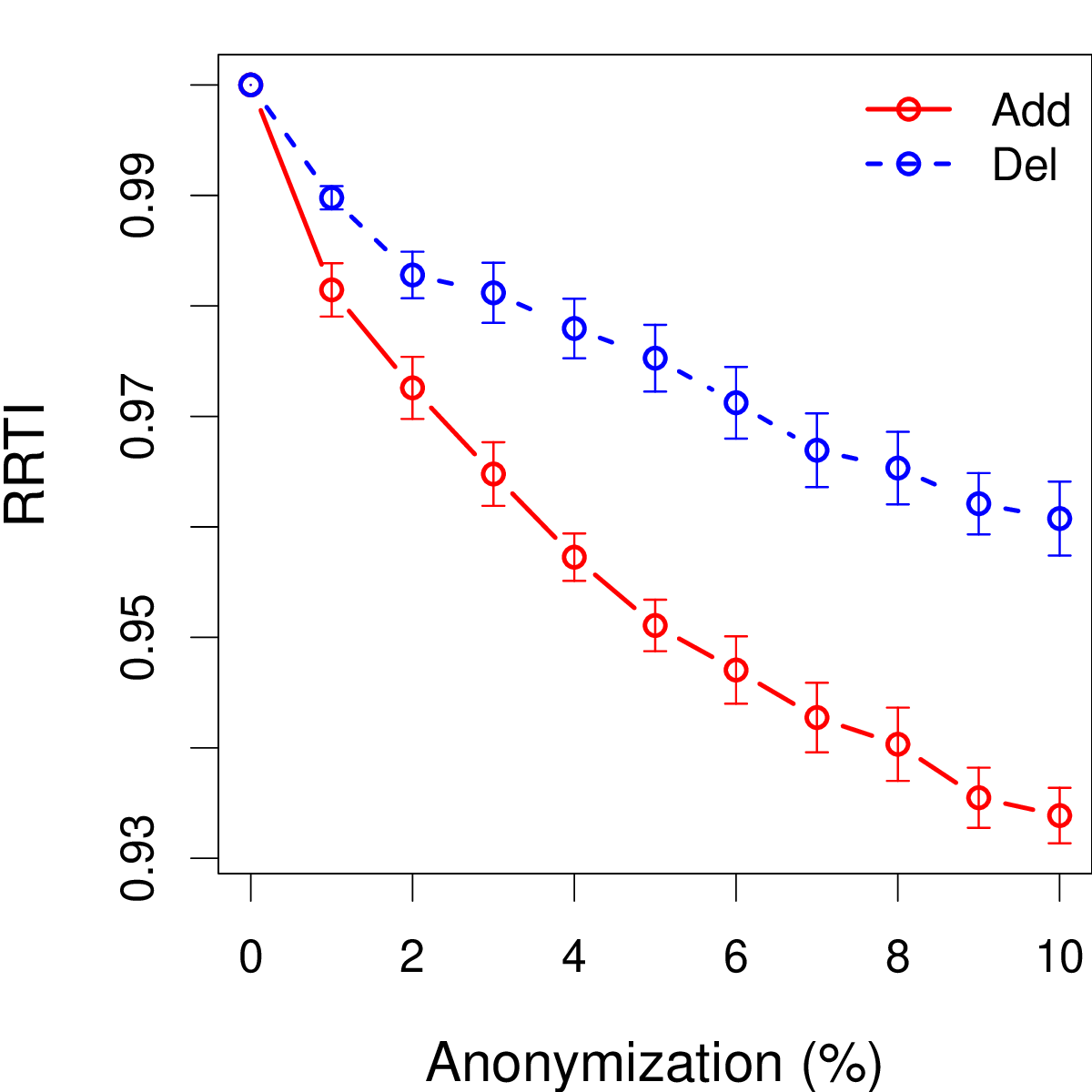}
  }
  \\
  \subfloat[Information flow]{
  	\label{fig:s3-FRV}
  	\includegraphics[width=0.23\linewidth]{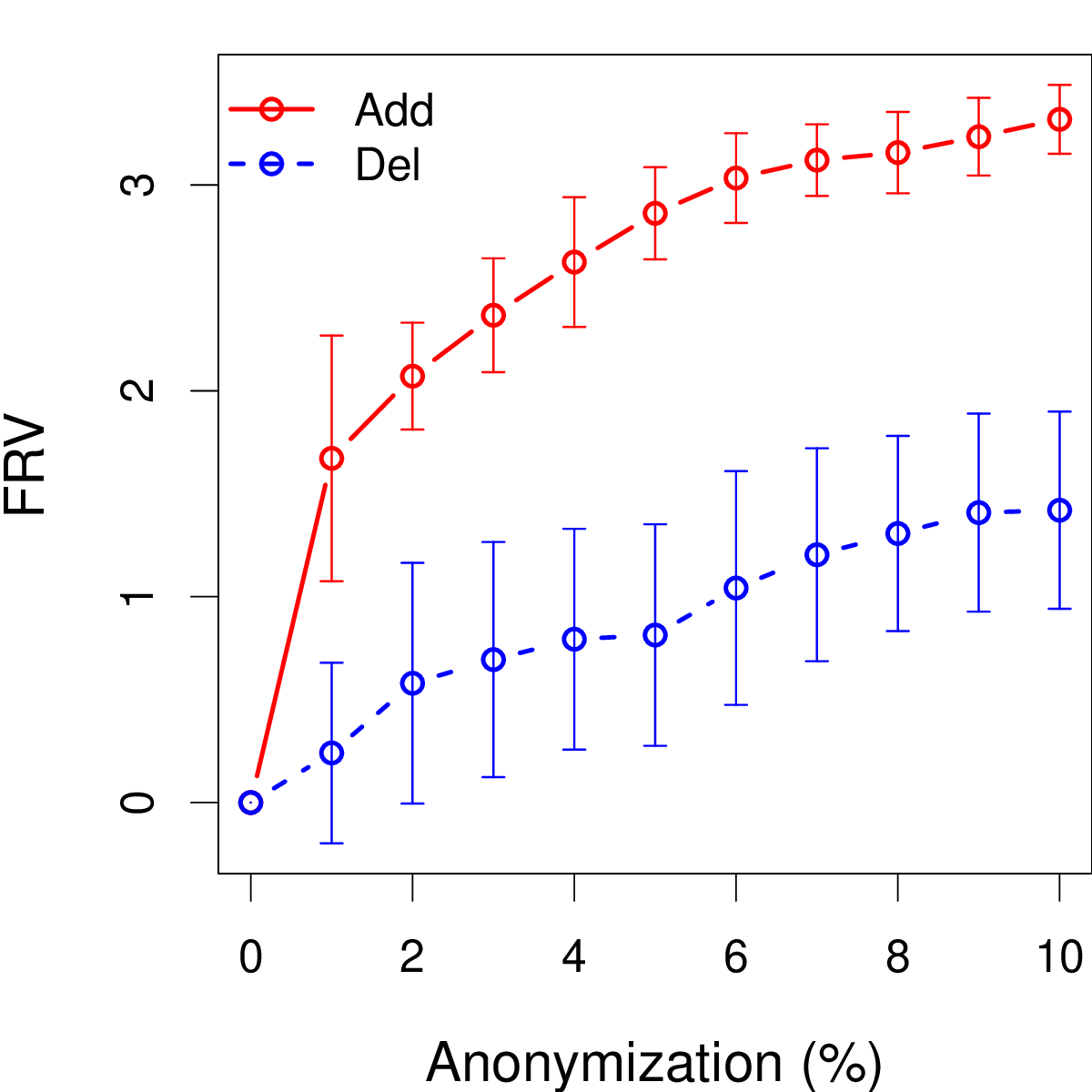}
  }
  ~
  \subfloat[$Cand_{\mathcal{H}_1}$ (Add)]{
  	\label{fig:sample-case-3_1-H1}
  	\includegraphics[width=0.23\linewidth]{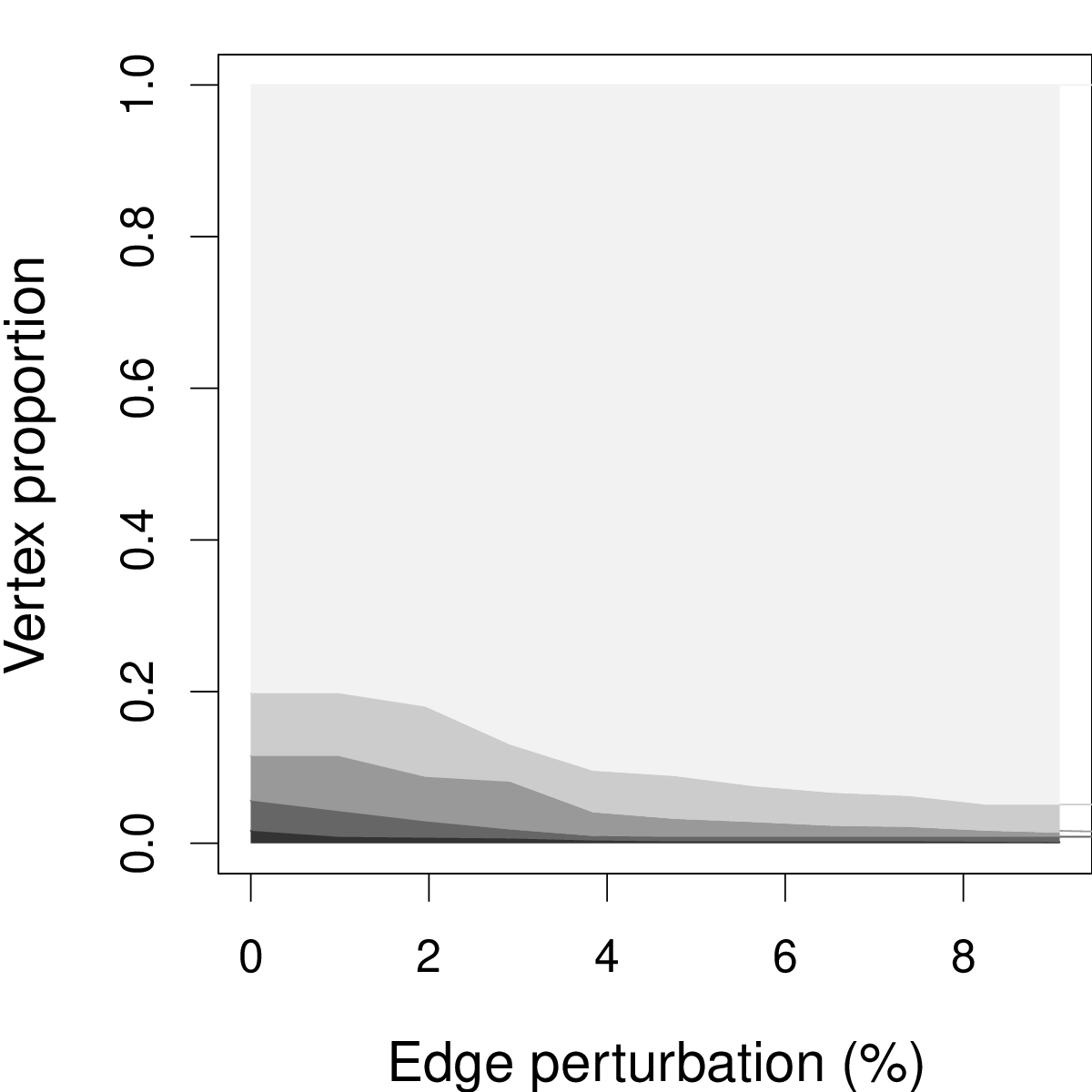}
  }
  ~
  \subfloat[$Cand_{\mathcal{H}_1}$ (Del)]{
  	\label{fig:sample-case-3_2-H1}
  	\includegraphics[width=0.23\linewidth]{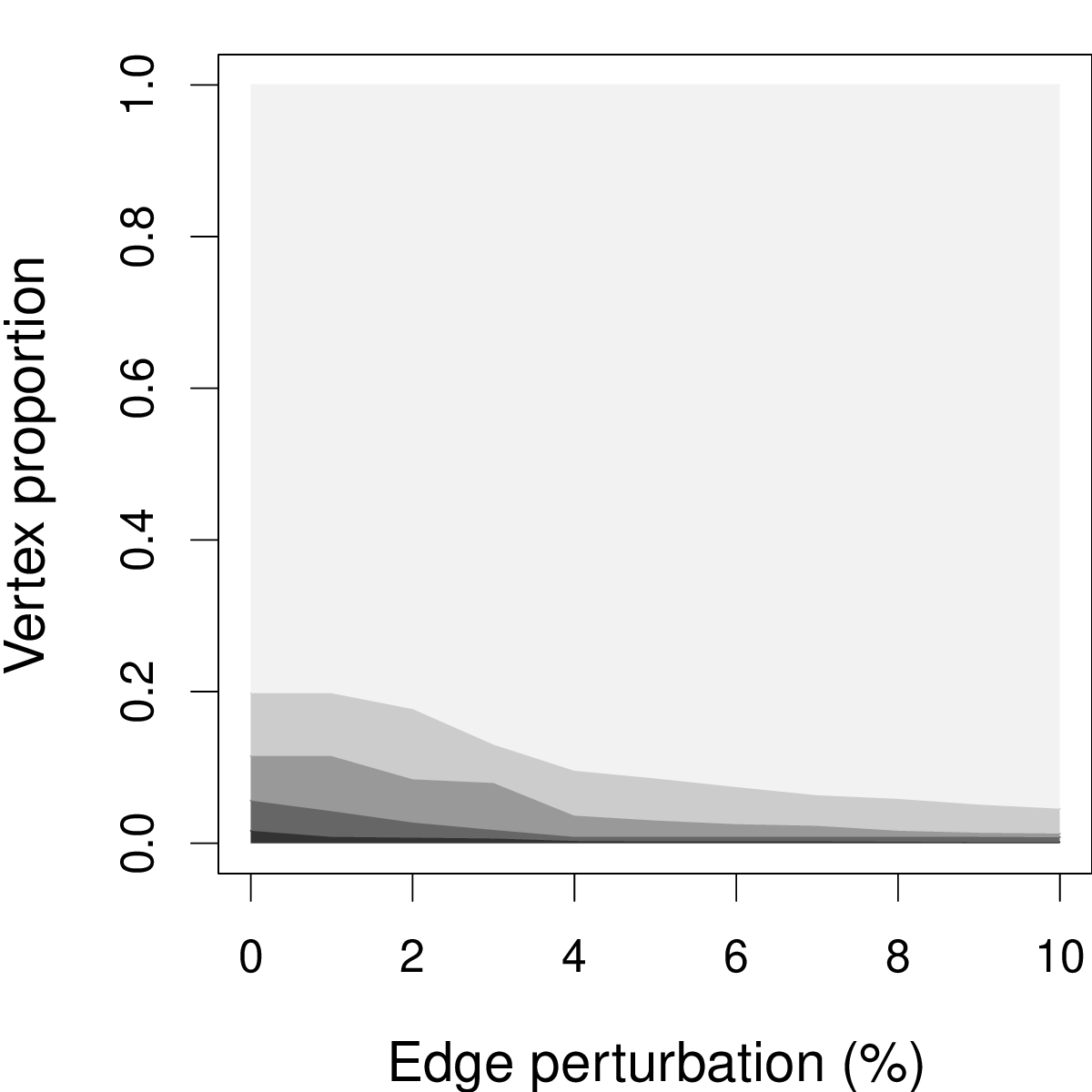}
  }
  ~
  \subfloat[1-Neighbourhood]{
  	\label{fig:s3-Neigh}
  	\includegraphics[width=0.23\linewidth]{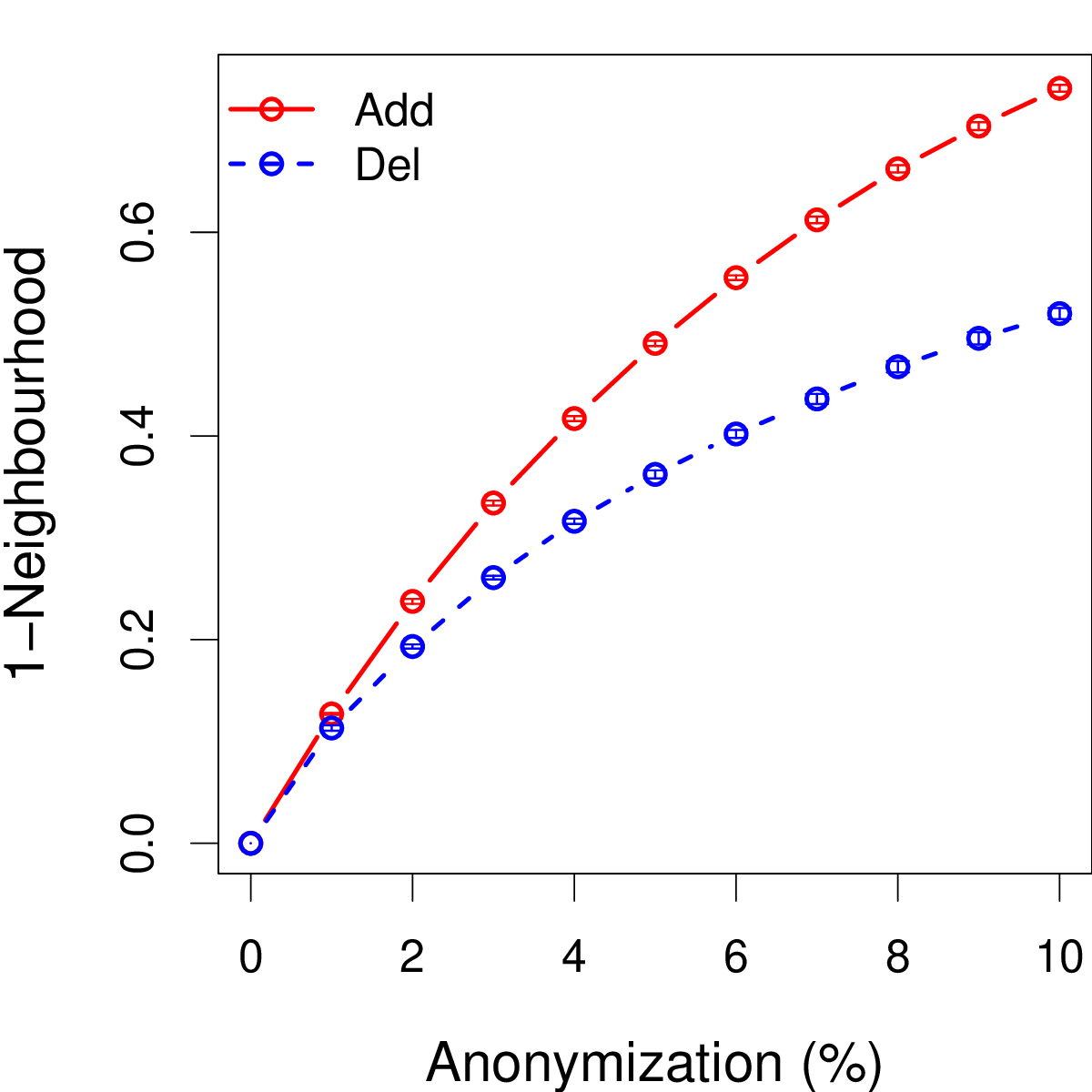}
  }
  \caption{Examples of our framework results for scenario III. The horizontal axis presents the anonymization (randomization \%), while vertical axis indicates the value of the original graph (leftmost point) and the evolution during anonymization processes.}
  \label{fig:s3}
\end{figure*}

An example using Hamsterster friendships dataset and two different randomization methods is presented in Figure \ref{fig:s3}. Anonymous graphs are generated applying edge addition (Add) and edge deletion (Del) at random over the whole edge set \cite{Casas-RomaEtAl:2016:AIRE}. We run 10 independent executions of each perturbation method to provide statistical significance since the perturbation process is stochastic.

For instance, the perturbation of average distance on anonymization percentage in range $[0, \ldots, 10\%]$ can be seen in Figure \ref{fig:s3-AD}, where we can see the mean value and the 95\% confidence interval error. It is clear to see how this metric evolves during the anonymization process. According to the figure, edge deletion better preserves the betweenness centrality in whole perturbation range. A similar behavior can be seen on betweenness centrality (Figure \ref{fig:s3-BC}), RRTI (Figure \ref{fig:s3-RRTI}) and FRV (Figure \ref{fig:s3-FRV}), where edge deletion considerably reduces the information loss on our specific information loss measures related to the most important users (hubs) and the information flow in the graph. However, taking into account the precision index using Walktrap algorithm, depicted in Figure \ref{fig:s3-WT}, we can see that edge addition clearly outperforms edge deletion in all ranges. Therefore, one could use this information to choose the perturbation method that better preserves the information loss depending on the subsequent graph mining task. 

Related to the re-identification and risk assessment analysis, we present the candidate set size results in Figures \ref{fig:sample-case-3_1-H1} and \ref{fig:sample-case-3_2-H1} for edge addition and deletion, respectively. As we can easily see, candidate set sizes shrink while perturbation increases on both edge perturbation techniques. The results are quite similar, since the range of edge perturbation is the same for both methods. Finally, the proportion of vertices that changed their 1-neighborhood during the anonymization process (Figure \ref{fig:s3-Neigh}) is considerably higher using the edge addition technique, which implies better protection from an attacker with knowledge based on the subgraph of distance 1 of the target vertices.

This framework aims to give information to researchers and practitioners to help them when choosing the algorithm and/or the parameters of the method itself. Since each perturbation method of algorithm has its strength and each dataset has its peculiarities, no method is the best on all situations. Thus, this framework can help us to choose the right algorithm and also its parameters to fulfill data utility and privacy constraints.

%% file: 9-conclusions.tex
\section{Conclusions}
\label{sec:conclusions}

In recent years several anonymization algorithms have appeared to protect users' privacy. However, it is quite difficult to compare data utility among them, since each work usually uses different measures to compute and evaluate information loss. In this paper we have proposed a framework to evaluate data utility and information loss on privacy-preserving graph data. We claim that some generic information loss measures can be used to compute and evaluate information loss. Nevertheless, metrics related to application-specific real-world problems must be defined and used to compute and compare data utility among methods and algorithms in literature. Additionally, our framework also considers some metrics to evaluate the re-identification and risk assessment considering an adversary with degree-based knowledge and also an adversary that has information about the 1-neighborhood of each target vertex.

We have presented three different scenarios to demonstrate the utility of this framework to compare different methods and parametrizations in order to achieve a good trade-off between privacy and data utility.

As a summary, we claim that our framework provides a standard way to compute both generic and specific information loss measures and can be easily used to perform comparisons among graph modification techniques, such as random-based, constrained-based and differential private methods. As we have proved through some application examples, this framework could help researchers and practitioners to choose the right parameters or algorithms to keep data utility and reduce information loss. Furthermore, it provides an easy and fair framework to compare different anonymization methods in order to quantify how the perturbation affects the graph structure and their application in some graph-mining tasks.

Many interesting directions for future research have been uncovered in this work. For instance, it could be interesting to extend the framework to include methods that change the vertex set. However, some generic information loss measures cannot be applied on graphs when the vertex set has changed. Furthermore, it would be thought-provoking to extend this analysis to other types of graph, such as directed or labelled graphs.